\documentclass[]{spie}

\newcommand{\minus}{\scalebox{0.75}[1.0]{$-$}}

\usepackage{natbib}

\usepackage{amsmath,amsfonts,amssymb}

\usepackage{amsmath,amsfonts,amssymb}
\usepackage{graphicx}
\usepackage[colorlinks=true, allcolors=blue]{hyperref}
\usepackage{titlesec}
\usepackage{xcolor}
\usepackage{array}
\usepackage{caption}
\usepackage{subcaption}
\usepackage[normalem]{ulem}

\title{Mitigating print-through effects through an optimized method for CFRP mirror production in Chile}

\author[a,b]{S. Castillo}
\author[c,d]{G. Hamilton}
\author[b,c]{N. Soto}
\author[a,b]{C. Lobos}
\author[b,c]{L. Pedrero}
\author[b,c]{C. Rozas}
\author[a,b,d]{A. Bayo}
\author[b]{P. Mardones}
\author[b,c,d]{H. Hakobyan}
\author[b,c,d]{C. Garc\'ia}
\author[b,c]{M.R. Schreiber}
\author[c,d]{W. Brooks}
\affil[a]{Instituto  de  F\'isica  y  Astronom\'ia,  Facultad  de  Ciencias,  Universidad de Valpara\'iso, Av. Gran Breta\~na 1111, 5030 Casilla, Valpara\'iso, Chile}
\affil[b]{N\'ucleo Milenio de Formaci\'on Planetaria - NPF, Valpara\'iso, Chile}
\affil[c]{Universidad T\'ecnica Federico Santa Mar\'ia}
\affil[d]{Centro Cient\'ifico Tecnol\'ogico de Valpara\'iso, CCTVal}

\titlespacing* 
    {\section} 
    {1em}
    {3.5ex plus 1ex minus .2ex}
    {2.3ex plus .2ex}

\authorinfo{Further author information: (Send correspondence to S.C)\\S.C: E-mail: scastillo@npf.cl, Telephone:+56982291703}

\begin{document} 
\maketitle

\begin{abstract}

In the manufacturing process of Carbon Fiber Reinforced Polymer (CFRP) mirrors (replicated from a mandrel) the orientation of the unidirectional carbon fiber layers (layup) has a direct influence on different aspects of the final product, like its general (large scale) shape and local deformations. In particular, optical methods used to evaluate the surface's quality, can reveal the presence of print-through, a very common issue in CFPR manufacture. In practical terms, the surface's irregularities induced, among other artifacts, by print-through, produce unwanted scattering effects, which are usually mitigated applying extra layers of different materials to the surface.
Since one of the main goals of CFPR mirrors is to decrease the final weight of the whole mirror system, adding more material goes in the opposite direction of that. For this reason a different layup method is being developed with the goal of decreasing print-through and improving sphericity while maintaining mechanical qualities and without the addition of extra material in the process.

\end{abstract}

\keywords{layup, cfrp, print-through, PFI, NPF}

\section{INTRODUCTION}
\label{sec:intro}  

The production of low cost segmented primaries for four to eight meters class infrared telescopes is one of the main technology requirements of the future infrared interferometers such as the Planet Formation Imager (PFI) \citep{Monnier18}. Developments in the manufacturing process of replicated carbon fiber reinforced polymers (CFRP) mirrors have been made by the collaboration of engineers, astronomers and experimental physicists from the N\'ucleo Milenio de Formaci\'on Planetaria (NPF) and the Centro Cient\'ifico Tecnol\'ogico de Valpara\'iso (CCTVAL).

The manufacturing of mirrors using composite materials has become a necessity for the next generation of optical elements of big size since it reduces weight, costs and time needed for production, while at the same time offering high  stiffness and low  thermal  expansion   coefficient \citep{wei17}. The method, called replication, consists of copying an optical quality mold surface with a series of layers of composite materials like carbon fiber impregnated with resin. This combination of materials makes it possible to clone the surface quality of the previously polished mandrel, but with the advantage of CFRP strength-to-weight ratio, which is better when compared to other commonly used materials, like glass or ceramic \citep{schmidt08}. Through this replication method, often called replica method, we found two main advantages: 
Using composite materials as carbon fiber reinforced polymer opens a range of possible uses when compared to traditional materials and secondly, the presence of a polymer allows the replica to match the surface quality of the mold down to nanometer scales.

Nevertheless, using composite materials such as CFRP comes with a price, which in this case is a very known issue called fiber print-through (FPT) \citep{hochhalter06}. In this paper we will be reporting on our experimental advances to optimize a layup method capable of mitigating the FPT (by changing the layup orientation) while keeping the elements that gives the CFRP replica its stiffness.

\section{MIRROR REPLICA METHOD}

\begin{figure}[!ht]
     \centering
     \begin{subfigure}[b]{0.45\textwidth}
       \centering
         \includegraphics[width=\textwidth]{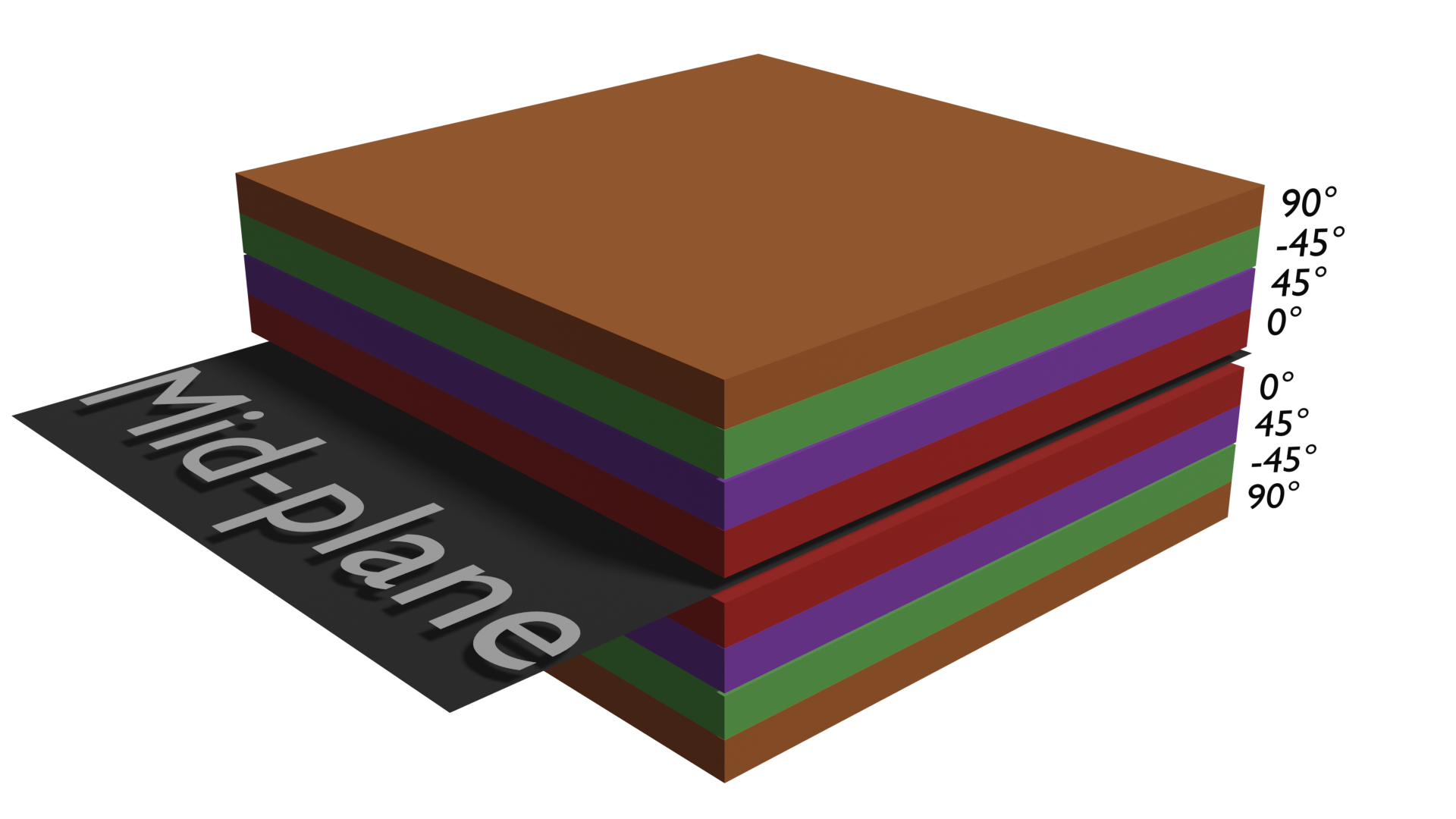}
         \caption{Symmetric laminate which consider an imaginary mid-plane}
         \label{fig:symmetric_laminate}
     \end{subfigure}
     \hfill
     \begin{subfigure}[b]{0.45\textwidth}
         \centering
         \includegraphics[width=\textwidth]{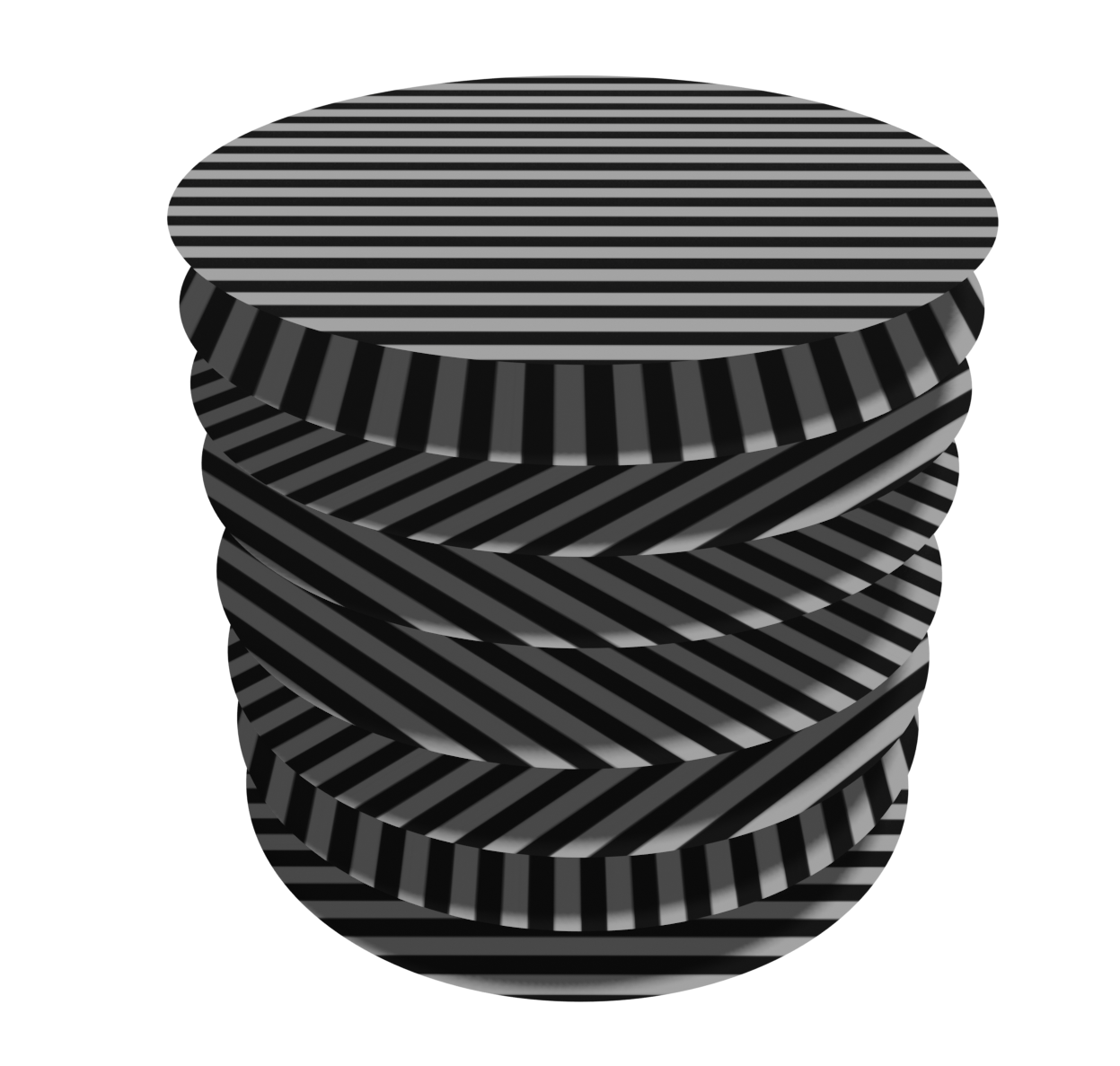}
         \caption{Quasi-isotropy laminate}
         \label{fig:Quasi-isotropy}
     \end{subfigure}
        \caption{Layup considerations}
        \label{fig:Layup consideration}
\end{figure}

In our experimental setup, the replica method consists mainly of a process called hand-layup, where consecutive layers of Unidirectional Carbon Fiber Reinforced Polymer (UD-CFRP) are positioned over the mold (mandrel) with specific orientations to achieve the desired mechanical properties. This layers can be made of different fabrics and different resins, but always a combination of both. Each one of this combinations have different thermo-mechanical properties and, from the mechanical point of view, the final layup has direct impact in those properties too \citep{HONGKARNJANAKUL2013549}. There are three important concepts when defining the layup for the specific properties of the material used: symmetry, balance and quasi-isotropy (see Fig.~\ref{fig:Layup consideration}).

A laminate is said symmetric when plies above a imaginary mid-plane are a mirror of those below the mid-plane \citep{joyce2003common}.

On the other hand, balance is achieved when an equal number of negatively and positively angled layers is employed, that usually translates to orthogonal pairs of layers  \citep{joyce2003common}.

Finally, quasi-isotropy aims to achieve an even distribution of the in-plane forces. Therefore, a quasi-isotropic laminate has either randomly oriented fiber in all directions, or has fibers oriented in a way that guarantees that an equal amount of strength is distributed all over the plane of the section. Generally, this can be achieved by using 4 orientations [0/90/45/-45] \citep{joyce2003common}.

\begin{figure}[!ht]
         \centering
         \includegraphics [width=\textwidth]{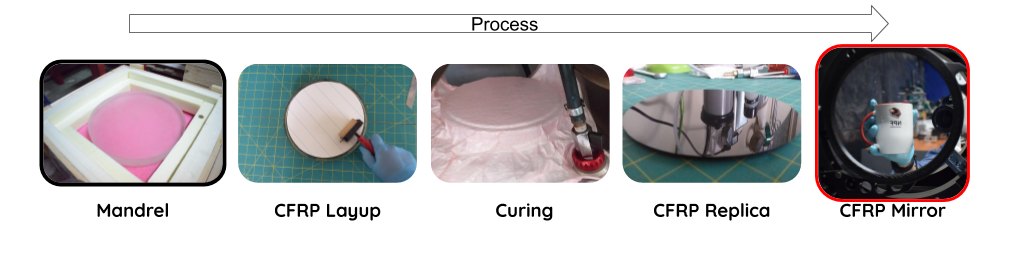}
         \caption{CFRP Mirror replication method}
         \label{fig:CFRP Mirror replication method}
\end{figure}

Once the proper layup is decided, a release agent is applied over the whole surface of the mold to avoid adhesion between the resin and the substrate. The hand-layup is then done and both CFRP and mold are enveloped in a series of materials which allow gases and resins to flow properly when cured. A brief summary of the process is presented below (and the full sequence of CFRP manufacture is displayed in Fig.~\ref{fig:CFRP Mirror replication method}).

To ensure proper vacuum, prior to the curing process, the vacuum level and the sealing quality must be checked to guarantee that there is no leaking. The last stage, curing, can be done in an autoclave or in an oven (often called out-of-autoclave) where heat is applied. The importance of this steps resides in the fact that the preimpregnated epoxy resin needs to be heated to crosslink \citep{pham05}, a process through which the resin acquires its crystal atomic structure, achieving its full mechanical properties. Besides that, while the resin is being heated it also becomes more fluid so it can fill any gaps between the layers and the substrate-CFRP interface. In addition, during this heating phase, gases can flow outside the laminate. The main advantage of using an autoclave is the higher pressure applied over the CFRP laminate, producing better quality pieces due to the smaller amount of bubbles allowed to form inside the laminate.

\section{CAUSES OF SURFACE ERRORS}

\begin{figure}[!ht]
     \centering
     \begin{subfigure}[b]{0.4\textwidth}
         \centering
         \includegraphics[width=\textwidth]{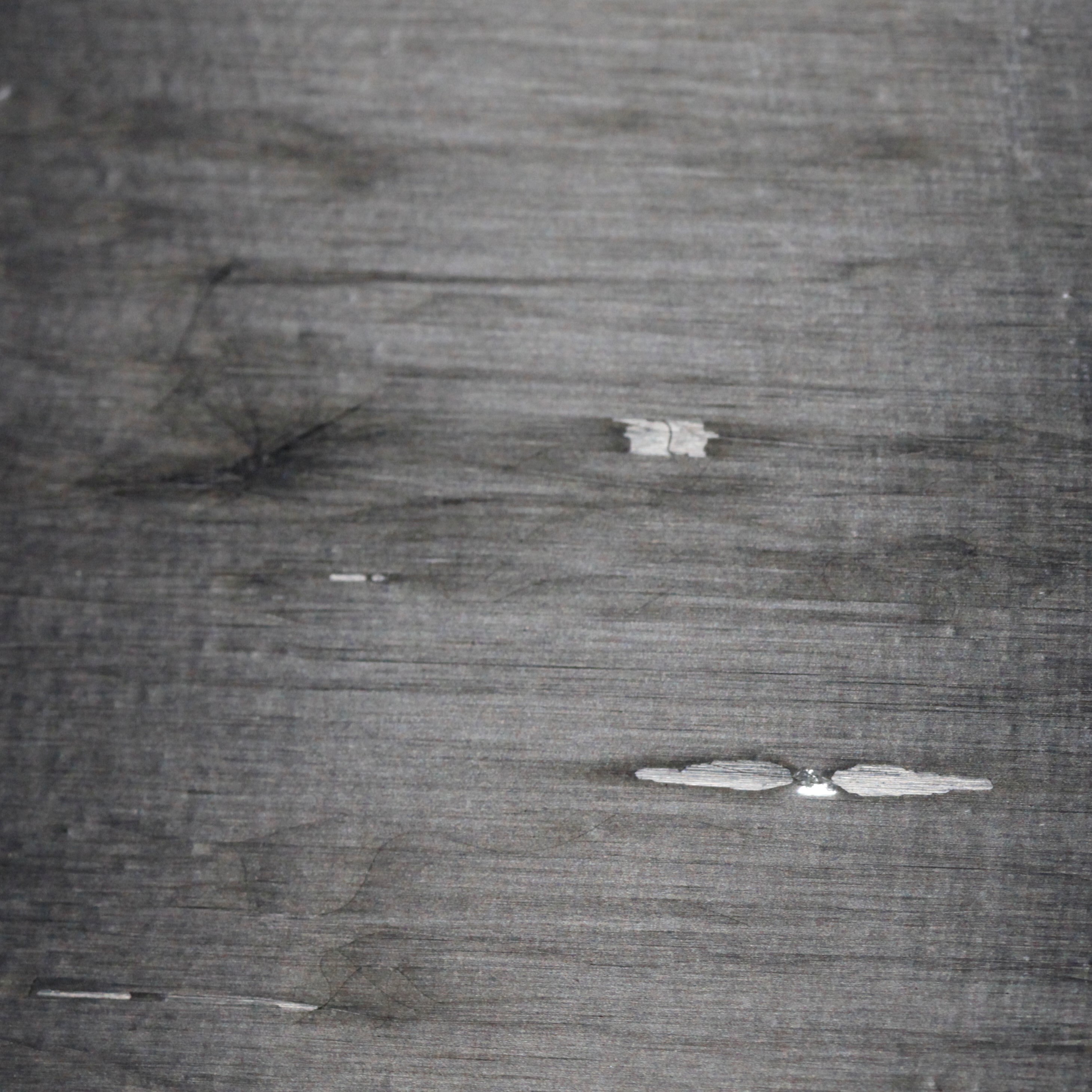}
         \caption{Surface contamination}
         \label{fig:surface}
     \end{subfigure}
     \hfill
     \begin{subfigure}[b]{0.4\textwidth}
         \centering
         \includegraphics[width=\textwidth]{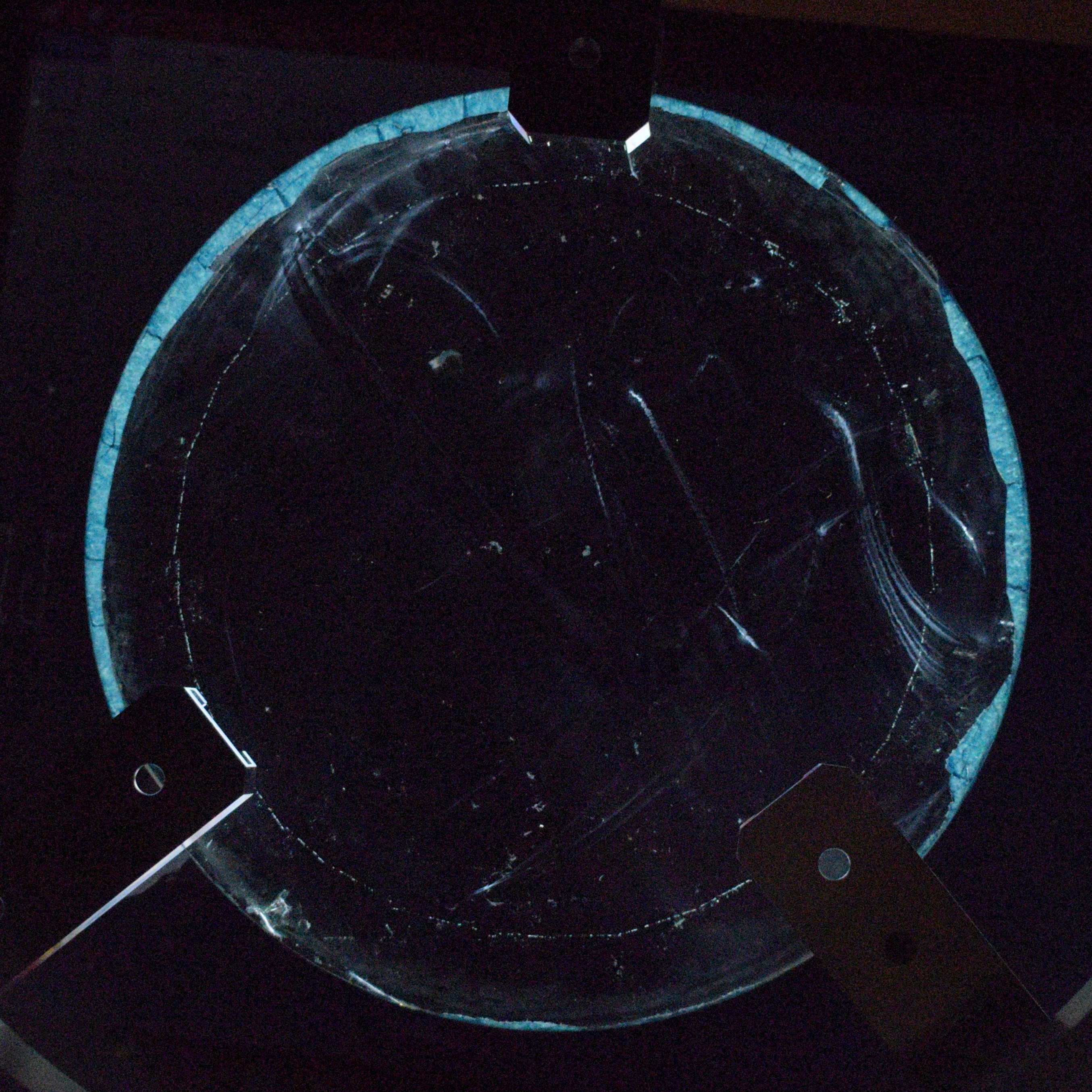}
         \caption{Mandrel defects}
         \label{fig:mandrel}
     \end{subfigure}
         \hfill
     \begin{subfigure}[b]{0.4\textwidth}
         \centering
         \includegraphics[width=\textwidth]{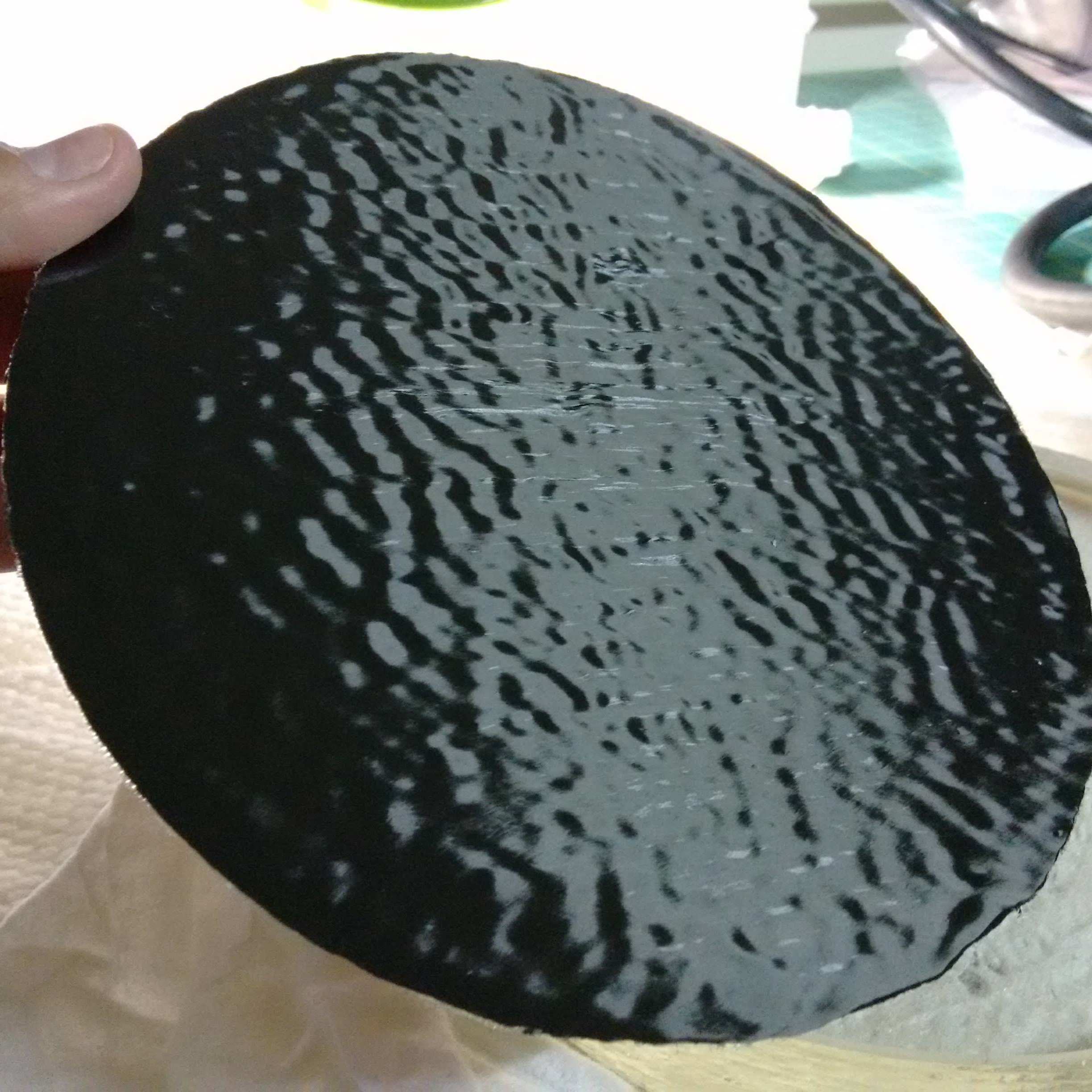}
         \caption{Bad curing cycle}
         \label{fig:curing}
     \end{subfigure}
    \hfill
     \begin{subfigure}[b]{0.4\textwidth}
         \centering
         \includegraphics[width=\textwidth]{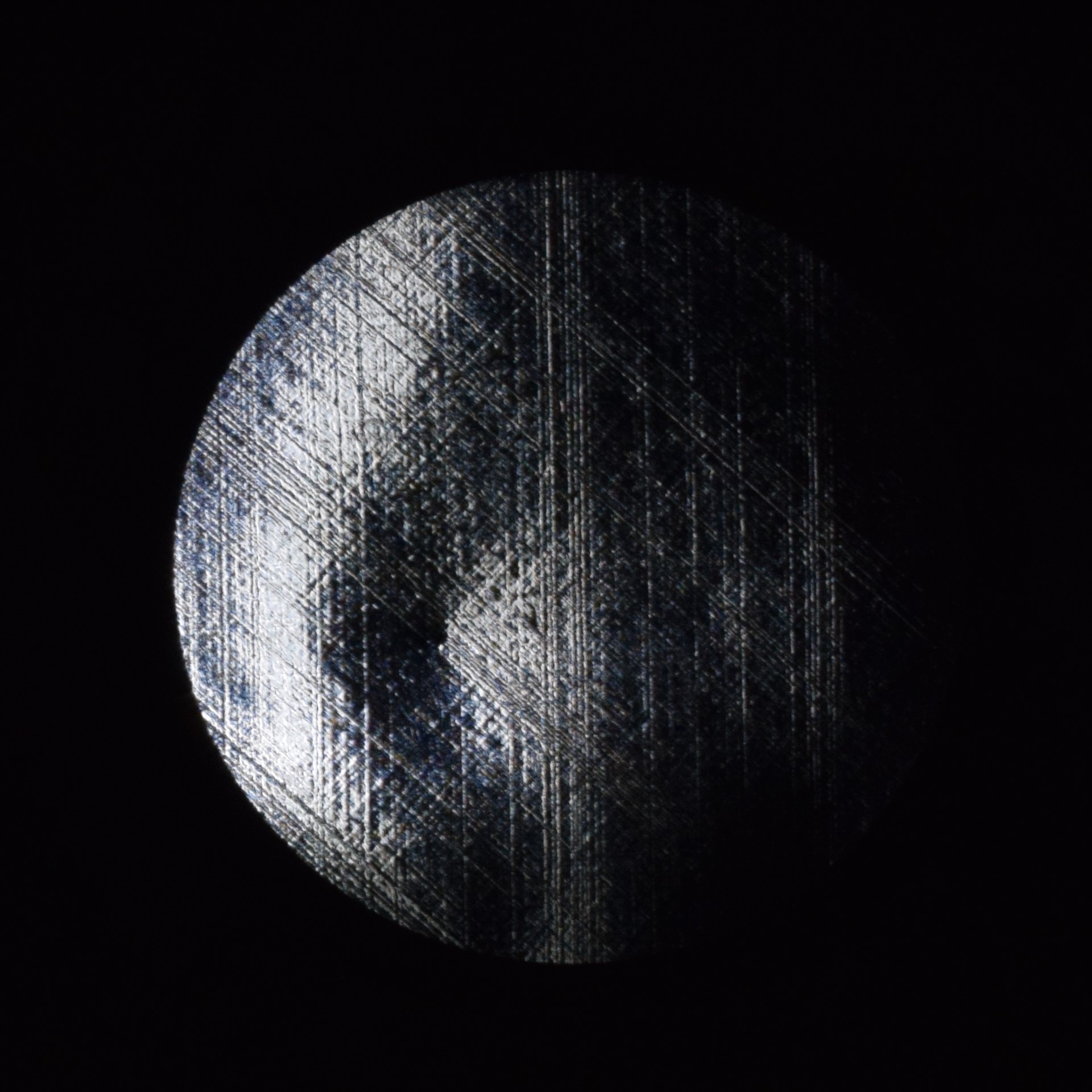}
         \caption{FPT example}
         \label{fig:fpt}
     \end{subfigure}
        \caption{Common surface errors}
        \label{fig:surface errors}
\end{figure}

During the replica process many factors can impact the final surface quality, such as surface contamination (Fig \ref{fig:surface}), bad layer compaction, the condition of the resin (Fig \ref{fig:curing}) and the layup itself (Fig \ref{fig:fpt}). 

Surface contamination can be mitigated by using clean rooms during the whole process, decreasing the amount of dust and other materials present in the air. 

On the other hand, despite having a controlled environment, leakage in the vacuum during the vacuum cycle can induce the presence of air and end up allowing bubbles between the different layers. These pockets of air will impact the accuracy of the final surface of the mirror. To expel the bubbles from the system it is necessary to apply enough pressure so they are forced out of the layers and a proper curing schedule needs to be set to ensure enough time for the resin to achieve lower viscosity. 

Finally, the layup itself plays a major role in the presence of surface aberrations which are mostly visible when layers are placed in orthogonal arrays where residual stress appears as ``fiber print-through" (FPT from now on) on the surface.

\section{OPTICAL ABERRATIONS}

\begin{figure}[!ht]
     \centering
     \begin{subfigure}[b]{0.475\textwidth}
         \centering
         \includegraphics[width=\textwidth]{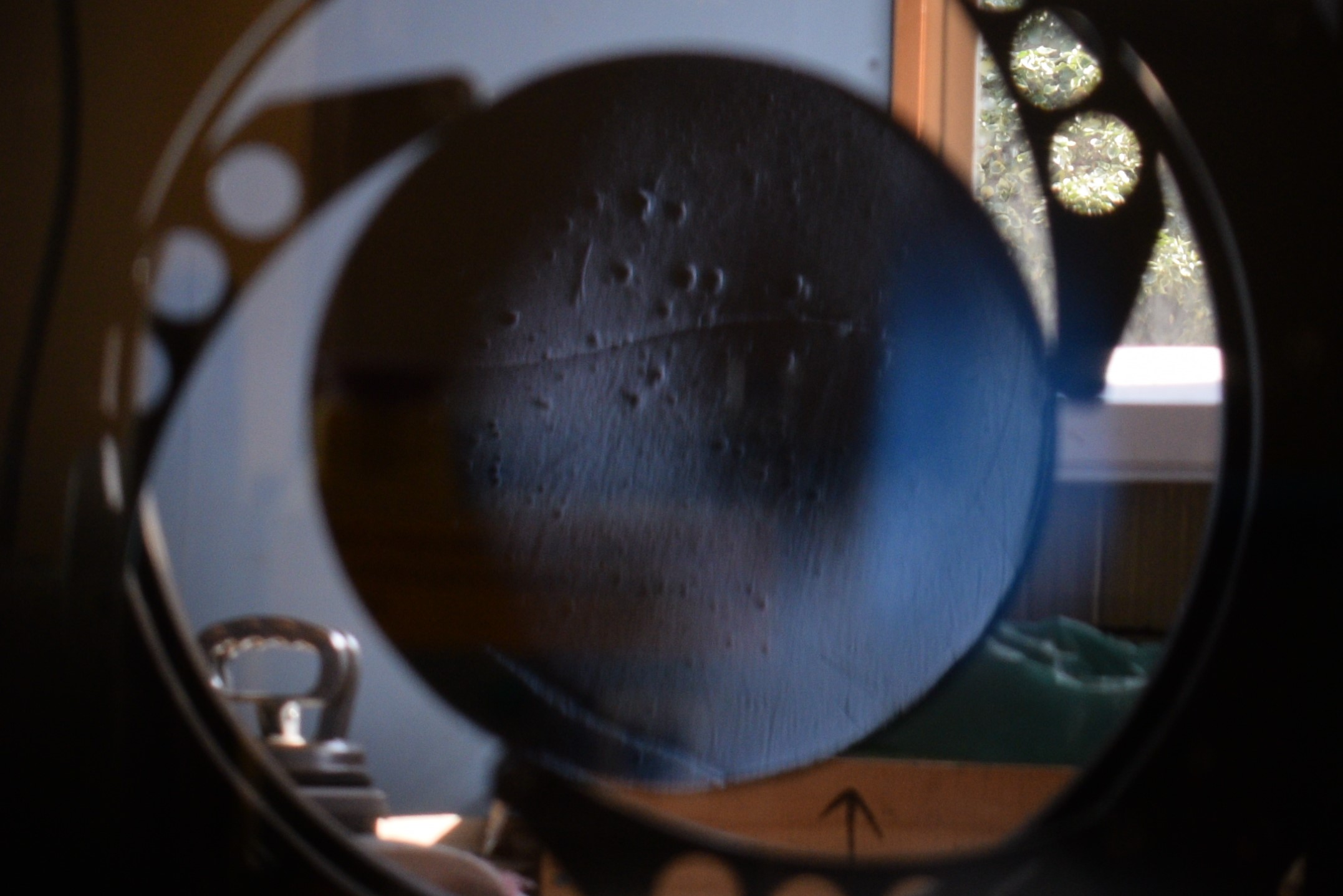}
         \caption{Focault test}
         \label{fig:focault}
     \end{subfigure}
     \hfill
     \begin{subfigure}[b]{0.475\textwidth}
         \centering
         \includegraphics[width=\textwidth]{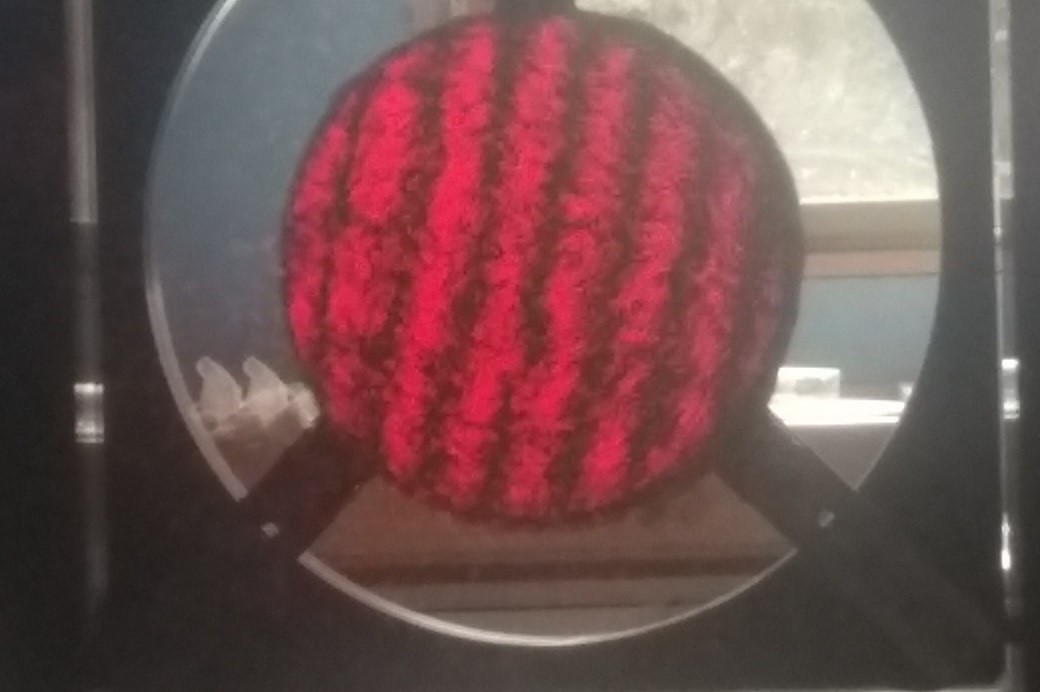}
         \caption{Interferometry test}
         \label{fig:interf}
     \end{subfigure}
        \caption{Optical tests}
        \label{fig:optical test}
\end{figure}

The traditional layup methods typically optimize the stress distribution to maximize the stiffness of the final product. However, for our application, surface quality is as important as mechanical properties, and the classical methods do not guarantee the former (see for example the optical tests displayed in Fig.~\ref{fig:optical test}). Therefore, new layup methodologies need to be explored and optimized.

The unidirectional CFRP is strong in the direction of the fiber and exerts zero force perpendicular to its orientation. In order to overcome this weakness the common way to proceed is to place a number of layers in different directions. With this approach FPT appears as a consequence (see Fig.~\ref{fig:interf}). One way to mitigate this effect is to apply extra resin over the surface after the first curing cycle. Even so, residual FPT can still be present and transferred to the extra resin layer surface. In this case, the amount of residual FPT is related to the thickness of this final resin layer and is also dependent of the laminate type: woven fibers show higher FPT in comparison to unidirectional fibers. This effect dominates the mid spatial frequencies and can be clearly seen in the surface waviness profiles.

\subsection{MEASURING METHODS}

To analyse this kind of aberration and understand its effects on the surface, visual deflectometry is used in order to estimate the surface shape and scattering, but this was not yet quantified in our experimental set-up because of a lack of instrumentation. A mechanical profilometer can also be used to measure mid spatial frequencies, where waviness can be observed. Finally, on the qualitative side of methodologies, naked eye tests are done such as focault, ronchi, interferometry and others (See Fig.~\ref{fig:optical test}).

\subsection{IMPROVING QUALITY SURFACE}

\subsubsection{ADDING A SOLID EXTRA LAYER OVER THE MIRROR AFTER FIRST CURE CYCLE}
\label{subsection:extra_resin}

As previously explained, due to the nature of the CFRP layers, the surface of the replica can present different types of aberrations, among them the FPT . With the goal of improving the surface quality of the mirrors, one or more extra resin layers are applied on the reflective face obtained after first curing cycle. Since this resin is a viscous liquid and can be easily spread over the surface, it is possible to replicate truthfully the quality of the mandrel's surface, like its roughness and waviness. Unfortunately, our experience shows that one single layer of resin is not enough to eliminate the FPT, as it was observed in our experiments. 

The surface quality measurements of the replicas before the first layer, after the first layer and after the second layer can be compared and analysed to help understand the efficiency of extra resin layer in FPT mitigation.

Even though our experience shows dramatic improvements after the second layer,there are problems that are intrinsic to the resin that suggest that extra-resin layering alone is not the definitive solution. 

One of the known problems has to do with the glass transition temperature (Tg) of the resin, which limits the range of applications that a CFRP mirror could have depending on how high the Tg is. This reason alone could lead to exclude space-based and direct sunlight applications. 

Besides, the epoxy resin suffers, like any other material, mechanical deformations, micro crackings and possible fractures due to humidity that can infiltrate in the micro crackings creating water deposits causing drastic degradation of this organic polymer. 

At last, but not less important, the average density of this resin is 1.4g/cm$^{3}$, similar to the carbon fiber density, but without the mechanical properties of the latter, therefore, arbitrary extra-layering could lead to an increase in fragility making the replicas more prone to failures while increasing the total weight of the system. All these considerations lead to the strong conclusion that the thickness of the extra resin layer is a key parameter to be considered and optimized. 

As a final aspect to be considered, the CTE mismatch between resin and carbon fiber could become an issue since it induces stress on the piece \citep{ahmed2012study}.

\subsubsection{SPUTTERING A METALLIC LAYER OVER THE MANDREL AND BONDING IT TO A PREVIOUSLY MADE CFRP MIRROR}

Another method proposed in the literature as a solution to the FPT problem is to sputter a series of different metals on the mandrel’s surface \citep{Steeves14}. In this framework, the surface of the mandrel has to be previously treated to prevent the adhesion of the metal to the glass while still allowing for exact surface cloning that can be later glued to an already prepared CFRP replica. Sputtering is a method for depositing a thin layer of atoms from a chosen material onto the surface of choice \citep{Behrisch82}. In this method, an electromagnetic field takes atoms of the material that will be deposited and orient these atoms gradually depositing them on the surface of choice. 

An advantage of this metallic multi-layer sputtering deposition is that these layers are effective at decreasing the FPT while also working as a reflective skin. So, in principle, this is a very promising double purpose method producing a final product that is closer to the goal, an already reflective mirror with better surface quality. However, the cost-effective scalability (in mirror size) of the process is far from trivial and the experiments shown in \emph{"Design, fabrication and testing of active carbon shell mirrors for space telescope applications"} \citep{Steeves14} were made only in 15cm hexagonal samples.

\section{PROPOSED METHOD}

\subsection{CHANGING THE ORIENTATION OF THE LAYERS}
\label{proposed method}

As previously mentioned, the traditional layup method optimizes the layers orientation with the goal of obtaining a mechanically resistant final piece, but it does not yield the optic quality needed. This work's proposition is to reorient a number of layers from the beginning to the end of the layup maintaining the basic considerations of balance, symmetry and quasi-isotropy, but allowing the mitigation of print-through thanks to the repetition of layers in the same direction.

\begin{figure}[h]
     \centering
         \centering
         \includegraphics[width=0.7\textwidth]{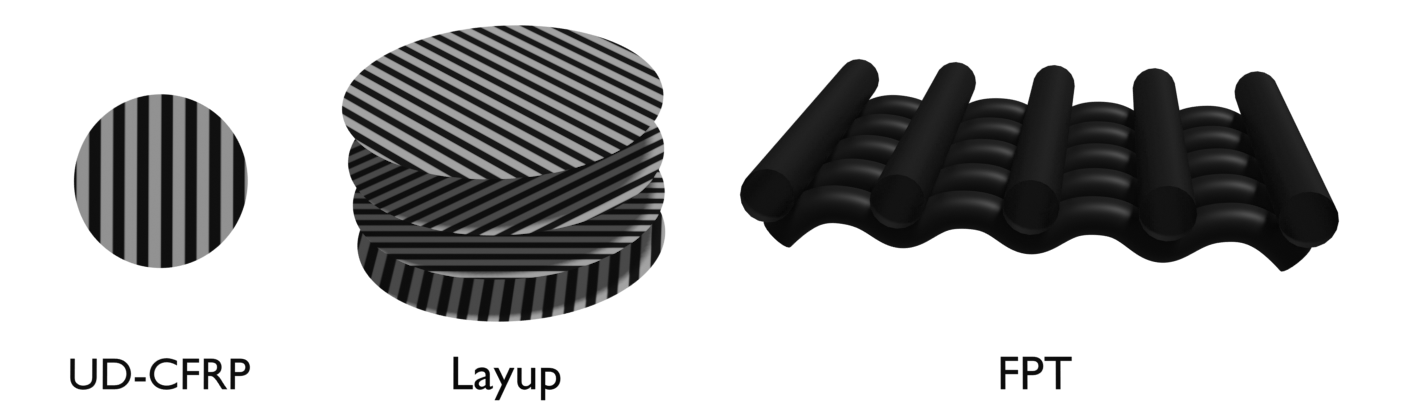}
         \caption{Representation of the effects of the micro tubes that compose CFRP in a traditional layup}
         \hfill
         \label{fig:layup_tradicional}
\end{figure}

If each layer is a pattern of unidirectional carbon fibers, we found that, in between different orientations, when more layers are stacked together in the same direction, the fibers are allowed to arrange themselves more efficiently decreasing the undulating surface profile. In the traditional method every layer has a different orientation as seen in Fig. \ref{fig:layup_tradicional} and the result is a surface with visible undulations coming from the inner layers. Our new method improves the initial replica's surface quality producing a final replica with better reflectivity since it decreases the FPT and the scattering effects, therefore having better roughness and waviness as seen in \ref{Tab:tests} and \ref{Tab:results}.

\begin{figure}[h]
     \centering
         \centering
         \includegraphics[width=0.7\textwidth]{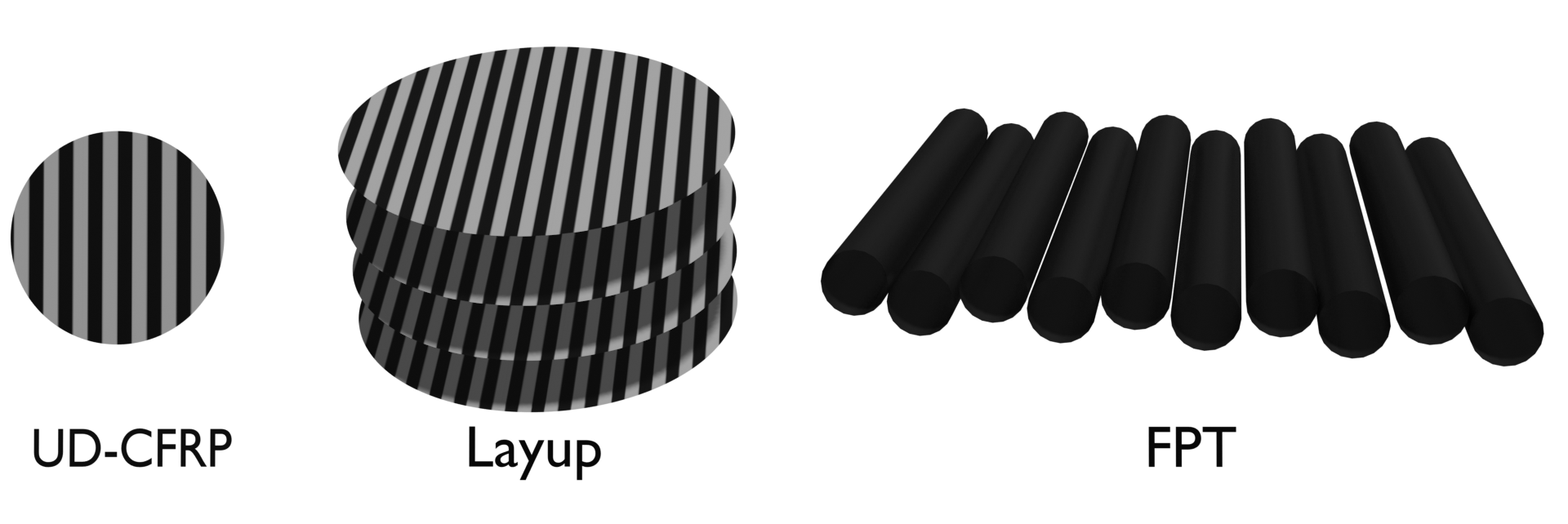}
         \hfill
         \caption{Representation of the effects of the micro tubes that compose CFRP in proposed layup}
         \label{fig:layup_propuesto}
\end{figure}

Extra resin layers are still going be necessary to suppress the remaining mostly-unidirectional FPT, but, with this layup method, its thickness can be reduced minimising the probability of mechanical failures due to the mismatch of the materials properties between the resin and the CFRP layers as  mentioned in subsection \ref{subsection:extra_resin}

In Table~\ref{Tab:tests} we present our results specifying, for each replica, its ID, mandrel, diameter, number of layers and layup used. Three of the replicas presented in the table were made with our proposed method (IDs 170-172-173), while the other four were made with traditional and non-traditional methods.

\begin{table}[ht]
\caption{Tests made}
\begin{center}
\begin{tabular}{ |c|c|c|c|c|c| } 
 \hline
 ID & Mandrel & Diameter(cm) & \# layers & Layup\\
    \hline
128 & CX200PYR & 19  & 24 & [0/90/15/\minus15/0/90]$_4$\\
   \hline
147 & CX200PYR & 19  & 8 & [0/10/20/30/40/50/60/70/80]\\
  \hline
157 & CX200PYR & 19  & 8 & [0/90/45/\minus45]s\\
  \hline
159 & CX200PYR & 19  & 8 & [0/90/45/\minus45]s\\
 \hline
170 & CX250CRO & 19  & 16 & [90$_4$/45/0/\minus45/0]s\\
 \hline
172 & CX250CRO & 19  & 24 & [90$_6$/0$_2$/45/\minus45/45/\minus45]s\\ 
 \hline
173 & CX500CRO & 48  & 16 & [90$_4$/45/0/\minus45/0]s\\
 \hline
 
\end{tabular}
\label{Tab:tests}
\end{center}
\end{table}

\newpage
\section{RESULTS}
\label{results}

\begin{figure}[!htt]
     \centering
     \begin{subfigure}[b]{0.475\textwidth}
         \centering
         \includegraphics[width=\textwidth]{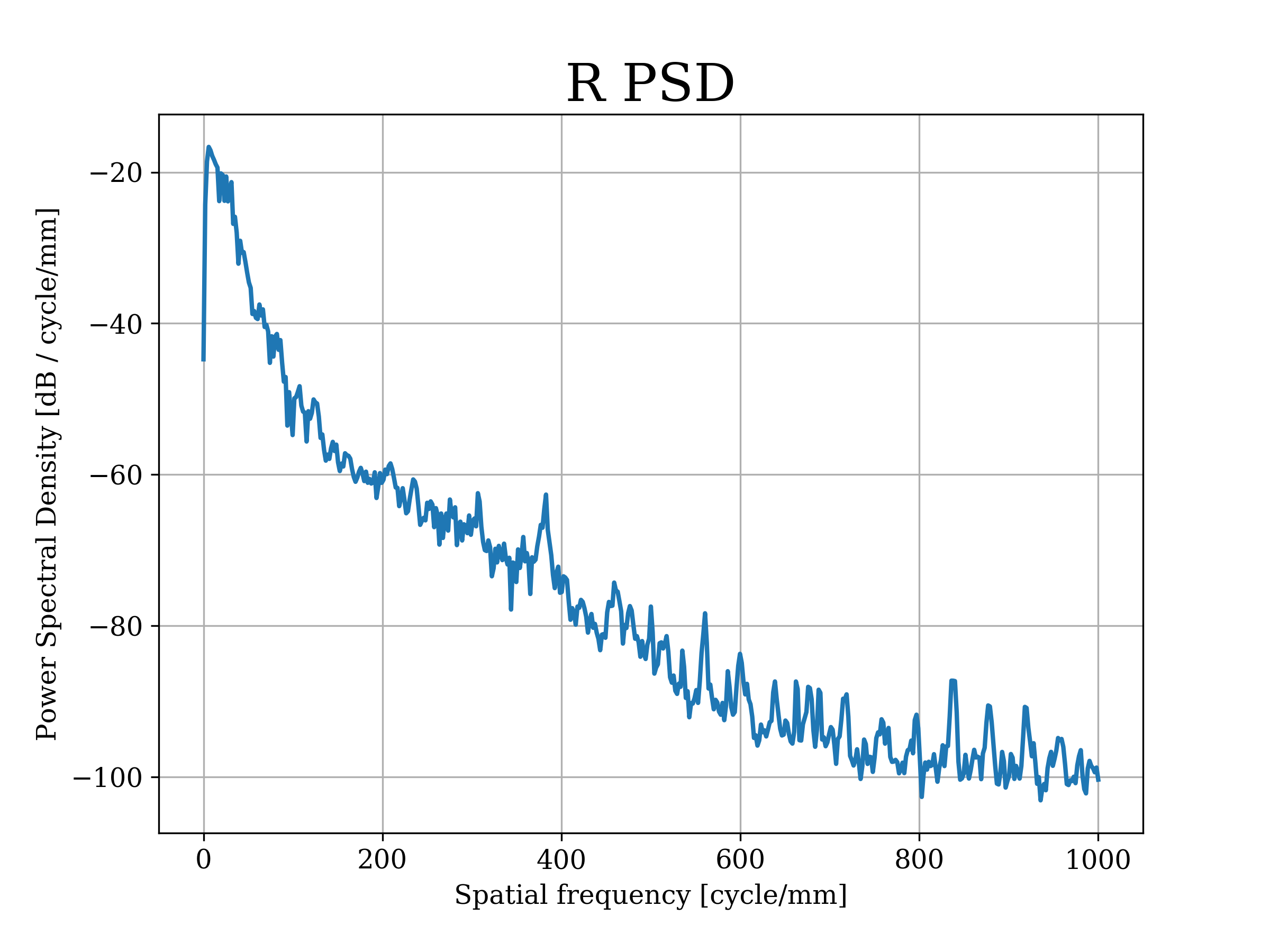}
         \caption{Rq Replica 157}
         \label{fig:157_r_master}
     \end{subfigure}
     \hfill
     \begin{subfigure}[b]{0.475\textwidth}
         \centering
         \includegraphics[width=\textwidth]{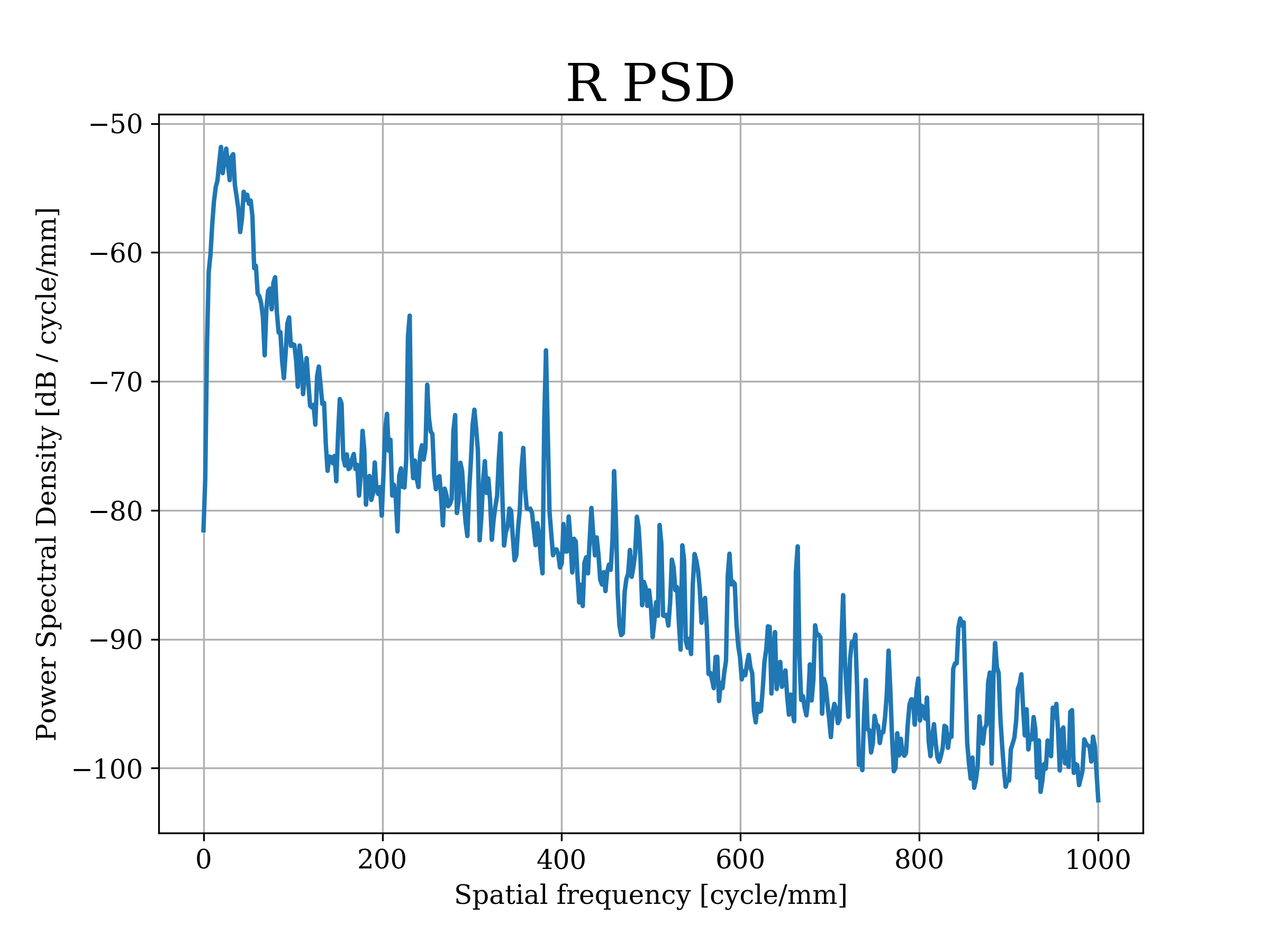}
         \caption{Rq Replica 170}
         \label{fig:170a_w_master}
     \end{subfigure}
        \caption{Rq Replica 157 and 170}
        \label{fig:rq_master}
\end{figure}

\begin{figure}[!ht]
     \centering
     \begin{subfigure}[b]{0.475\textwidth}
         \centering
         \includegraphics[width=\textwidth]{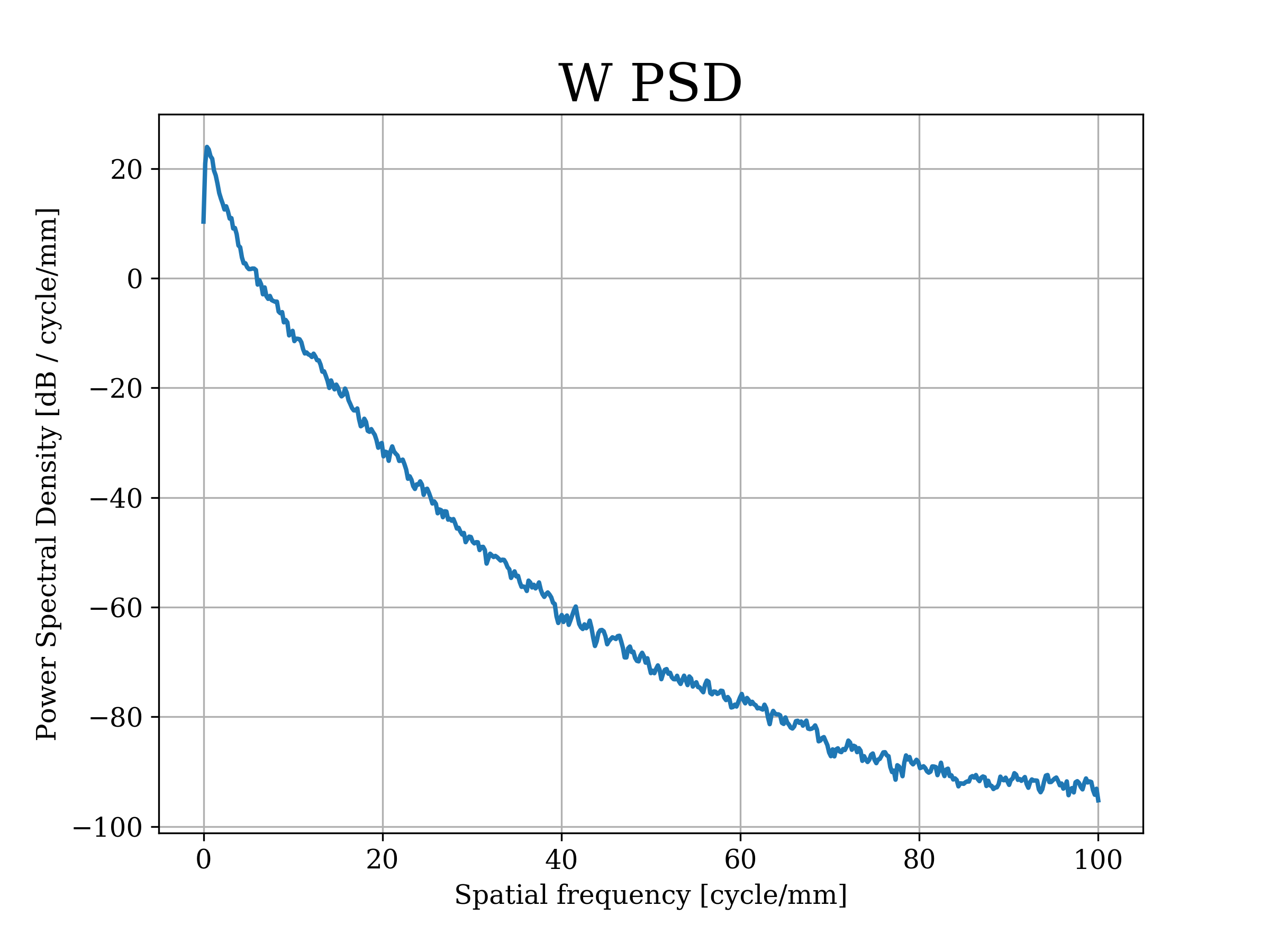}
         \caption{Wq Replica 157}
         \label{fig:157_w_master}
     \end{subfigure}
     \hfill
     \begin{subfigure}[b]{0.475\textwidth}
         \centering
         \includegraphics[width=\textwidth]{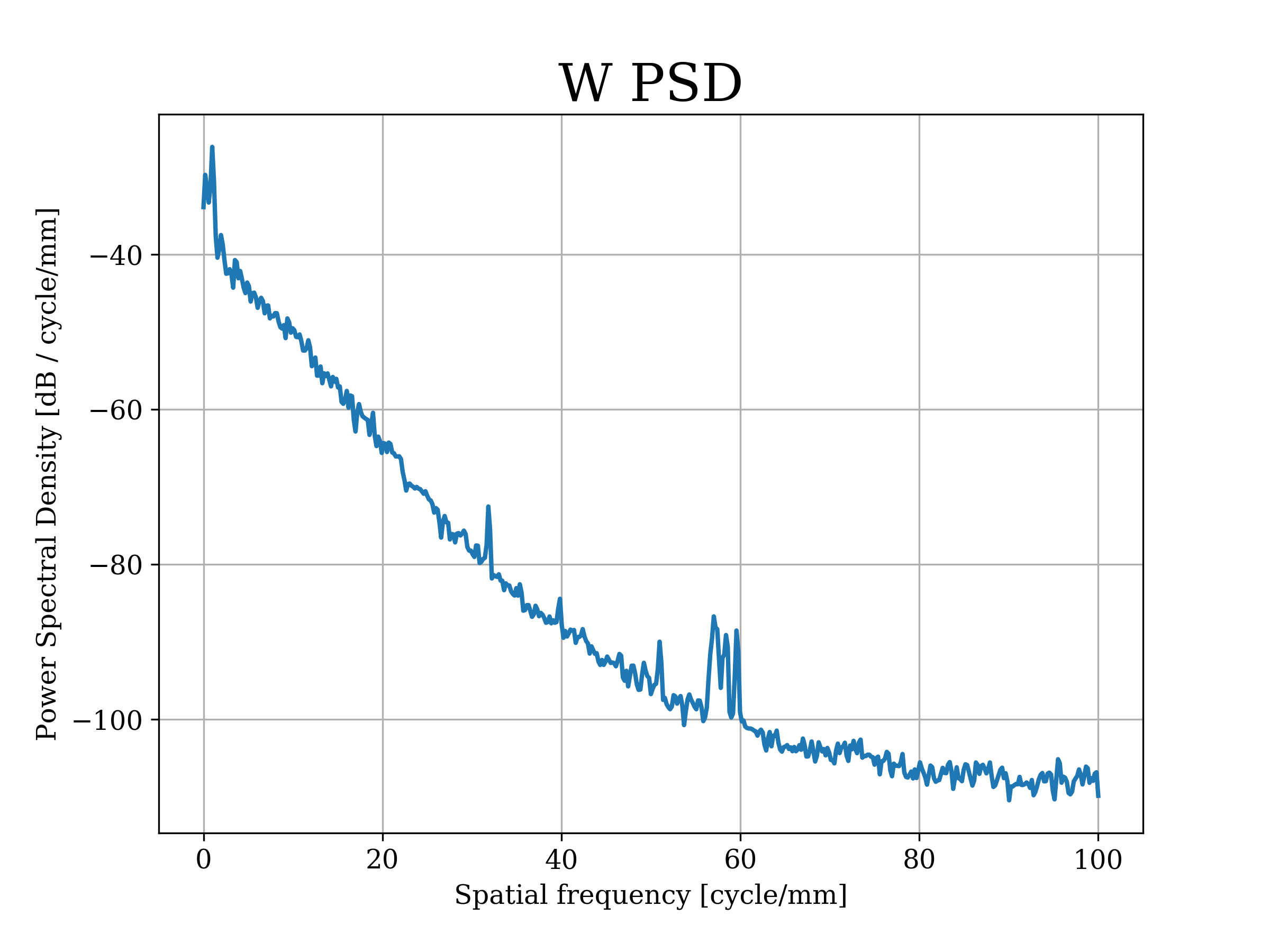}
         \caption{Wq Replica 170}
         \label{fig:170b_w_master}
     \end{subfigure}
        \caption{Wq Replica 157 and 170}
        \label{fig:Wq_master}
\end{figure}

Analyzing these changes in layup orientation we have observed improvements both in roughness and waviness, measuring around four times better quality using our method, mostly in 19 cm spherical CFRP replicated mirrors. Regardless, further testing is needed in terms of scalability of the replica size. In Fig.~\ref{fig:rq_master} and Fig.~\ref{fig:Wq_master} we present the results for Rq and Wq for replicas with the traditional method versus the proposed method.
   
\newpage
As it can clearly be seen in Table~\ref{Tab:results}, replicas made with our proposed method have better roughness, and even better waviness, on the first curing cycle. This could lead to thinner extra layers to mitigate FPT, yielding a lighter final piece when compared to CFRP mirrors made with the traditional layup.

\begin{table}[ht]
\caption{Results}
\begin{center}
\begin{tabular}{ |c|c|c| } 
 \hline
 ID & Rq & Wq\\
   \hline
128 & 17 nm & 152 nm  \\
  \hline
147 & 235 nm & 2896 nm  \\
  \hline
157 & 539 nm & 17577 nm  \\
  \hline
159 & 169 nm & 10225 nm  \\
 \hline
170 & 12 nm & 51 nm  \\
 \hline
172 & 15 nm & 41 nm \\ 
 \hline
173 & 133 nm & 658 nm \\ 
 \hline
 
 \end{tabular}
\label{Tab:results}
\end{center}
\end{table}

From Table \ref{Tab:results} we can also see the improvement even when compared to a 24 layers CFRP mirror (ID 128), meaning that with less layers we can achieve better initial surface quality, specially at mid spatial frequencies.

\section{CONCLUSIONS}

With our experiments we have shown that non-traditional layup procedures can help mitigate critical aspects of CFRP replicas such as FPT, giving us an insight of how crucial this consideration could be on the next generation of lightweight astronomical mirrors. As future work we will systematically characterize the relation between the number of CFRP layers in the same direction and the amount of mitigated FPT. 

The knowledge of how these parameters are related will allow us to determine the optimum setup for a specific replica considering the total number of layers, diameter, the curing technique and others.

\appendix 
\section{Azimutal profile for CFRP mirrors}

In this appendix we present the azimutal profiles obtained for every replica previously discussed both in \ref{proposed method} and \ref{results} that were not included in the body of the paper to improve readability. All the profiles here presented were obtained with a portable surface roughness tester Mitutoyo SJ-410.

\begin{figure}[!ht]
     \centering
     \begin{subfigure}[b]{0.45\textwidth}
         \centering
         \includegraphics[width=\textwidth]{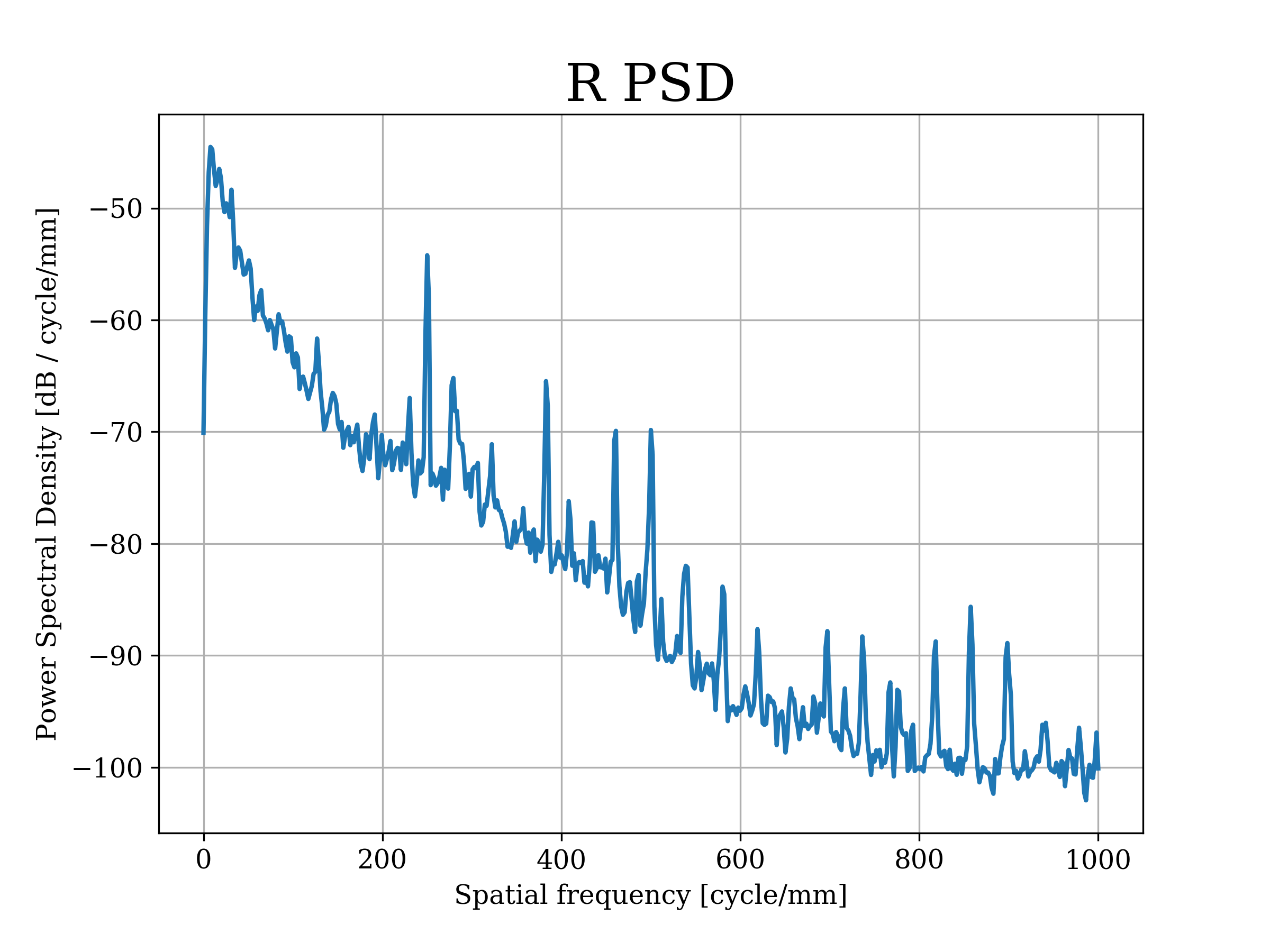}
         \caption{Rq Replica 128}
         \label{fig:128_r_master}
     \end{subfigure}
     \hfill
     \begin{subfigure}[b]{0.45\textwidth}
         \centering
         \includegraphics[width=\textwidth]{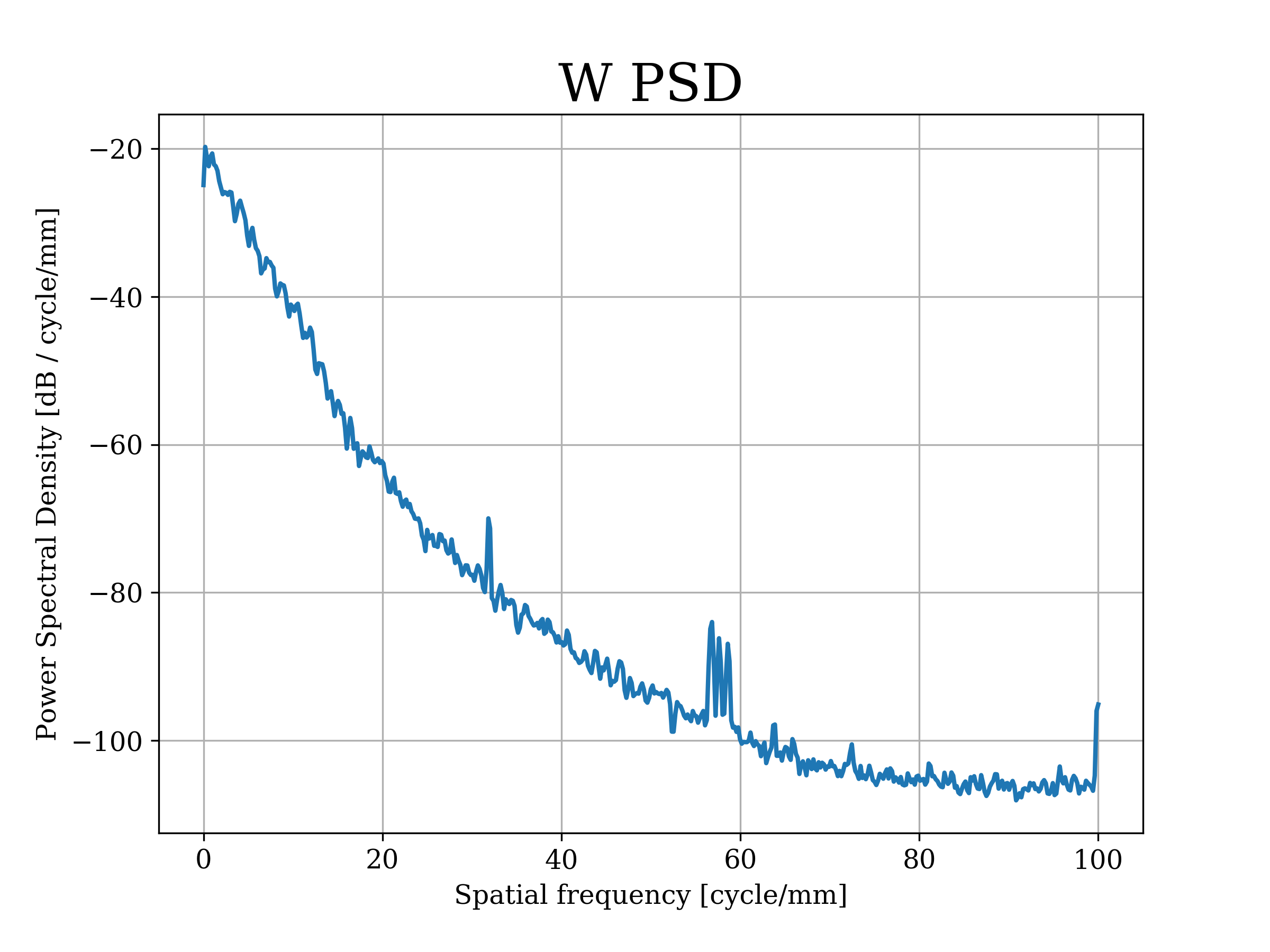}
         \caption{Wq Replica 128}
         \label{fig:128_w_master}
     \end{subfigure}
        \caption{Replica 128 Rq and Wq}
        \label{fig:128_rq_wq_master}
\end{figure}

\begin{figure}[!ht]
     \centering
     \begin{subfigure}[b]{0.45\textwidth}
         \centering
         \includegraphics[width=\textwidth]{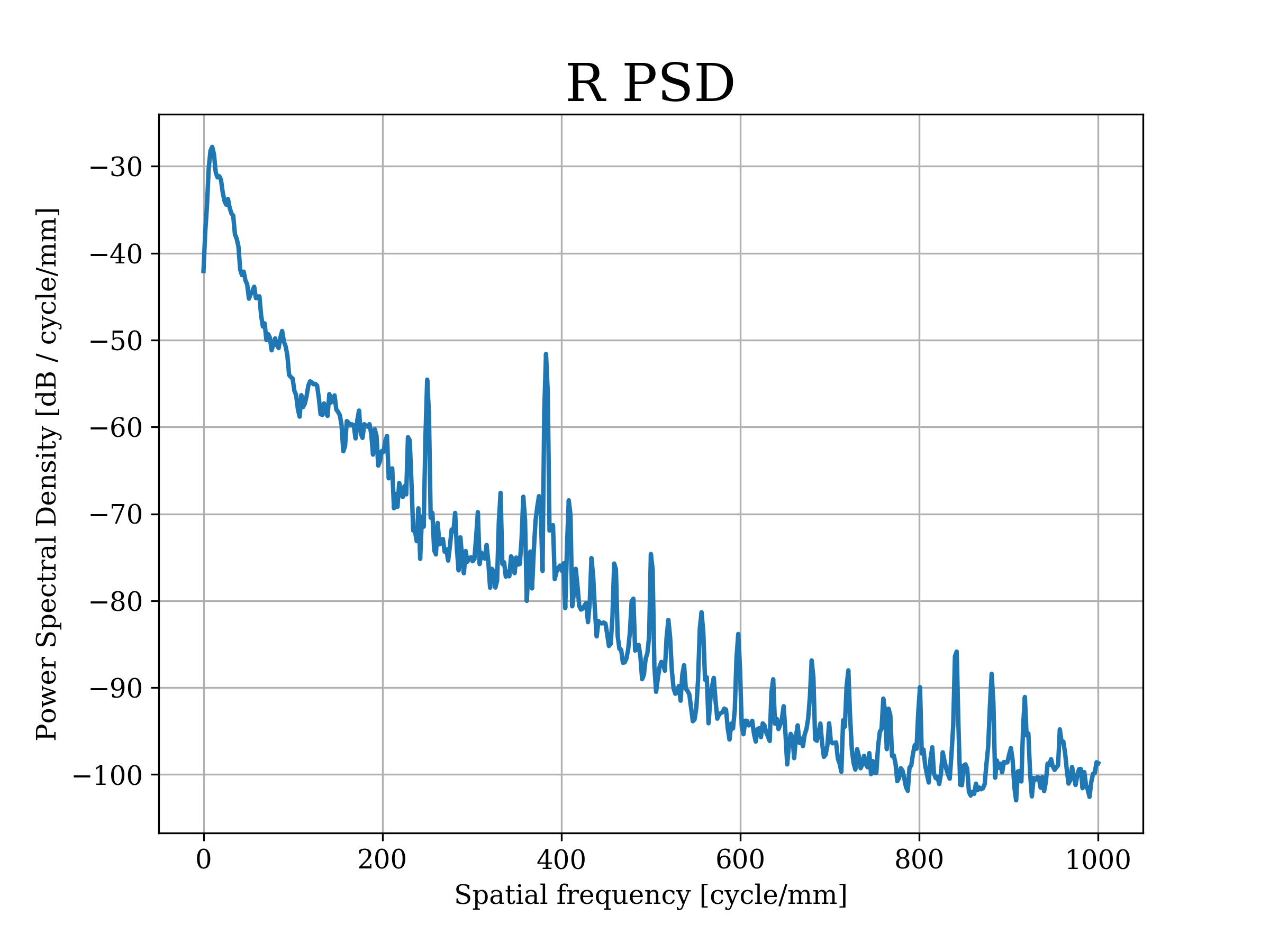}
         \caption{Rq Replica 147}
         \label{fig:147_r_master}
     \end{subfigure}
     \hfill
     \begin{subfigure}[b]{0.45\textwidth}
         \centering
         \includegraphics[width=\textwidth]{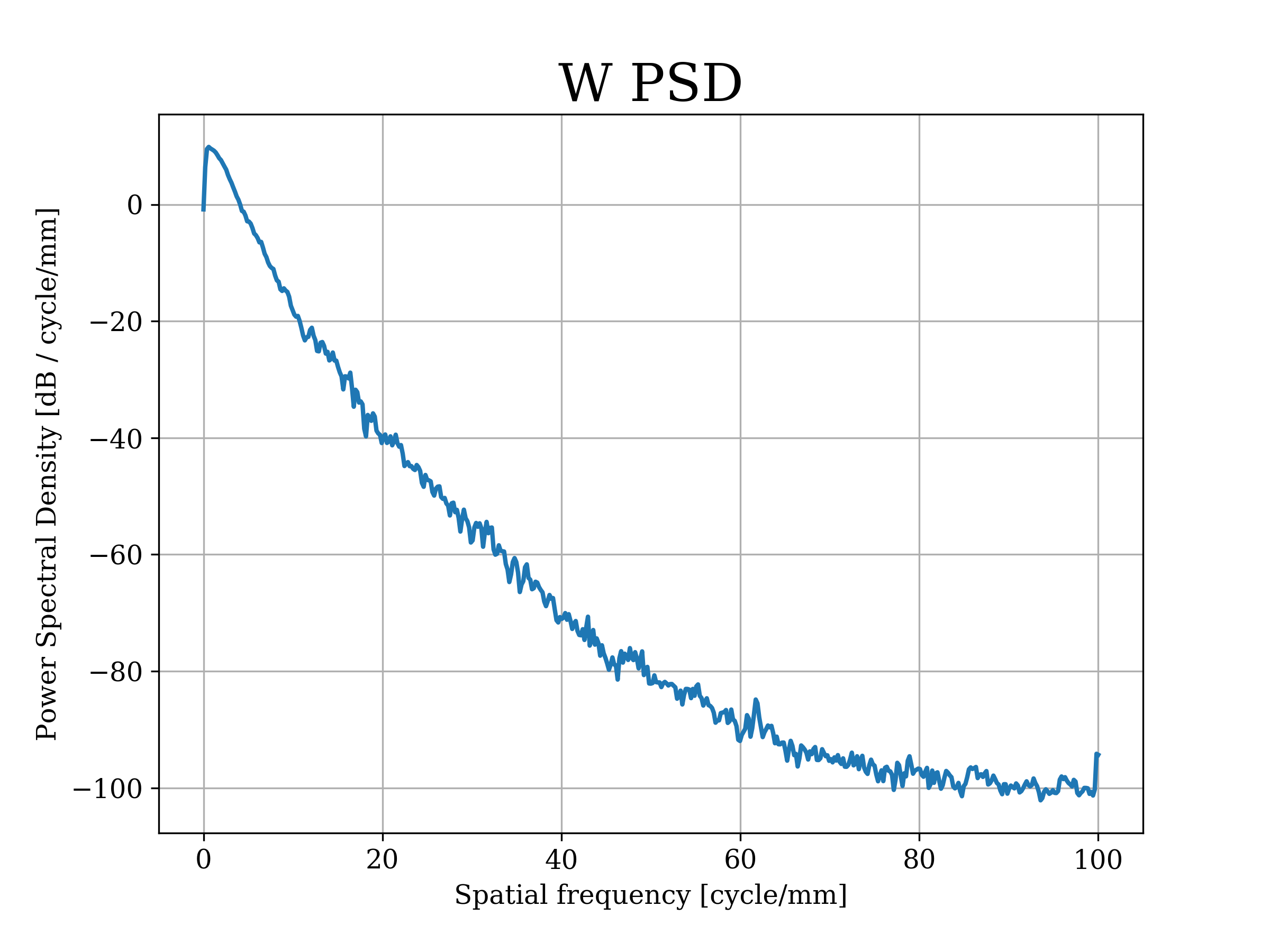}
         \caption{Wq Replica 147}
         \label{fig:147_w_master}
     \end{subfigure}
        \caption{Replica 147 Rq and Wq}
        \label{fig:147_rq_wq_master}
\end{figure}

\begin{figure}[!ht]
     \centering
     \begin{subfigure}[b]{0.45\textwidth}
         \centering
         \includegraphics[width=\textwidth]{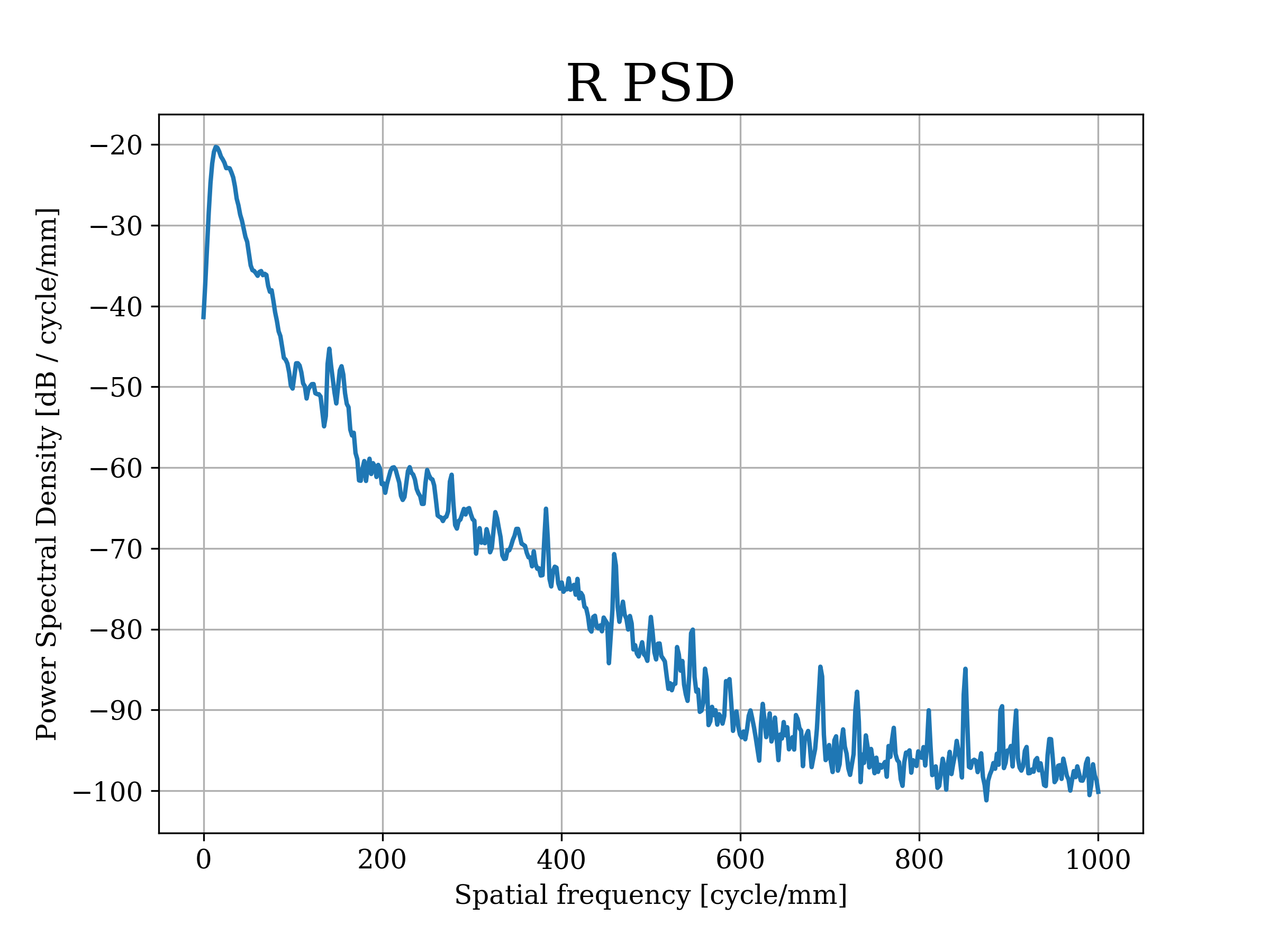}
         \caption{Rq Replica 159}
         \label{fig:159_r_master}
     \end{subfigure}
     \hfill
     \begin{subfigure}[b]{0.45\textwidth}
         \centering
         \includegraphics[width=\textwidth]{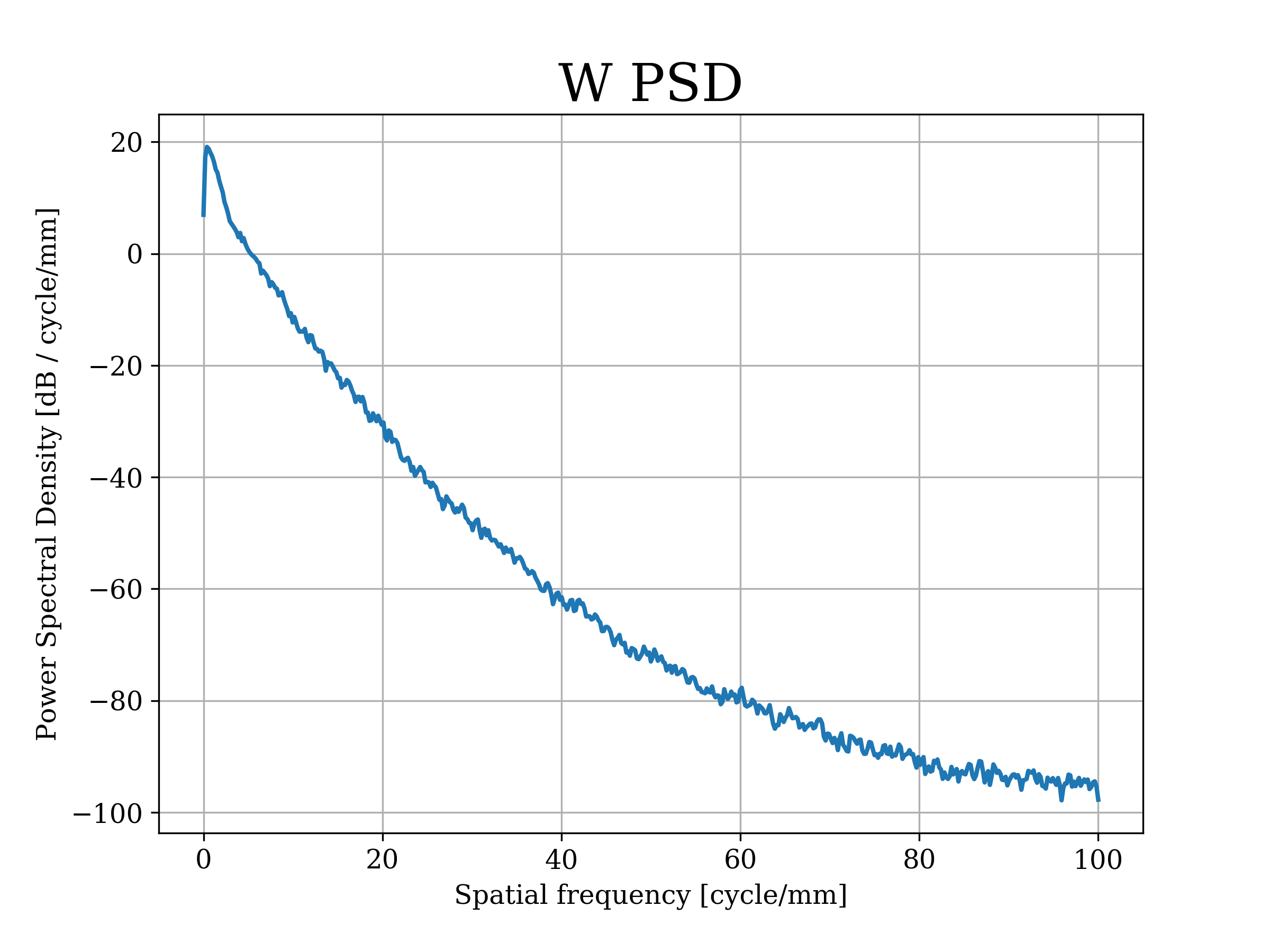}
         \caption{Wq Replica 159}
         \label{fig:159_w_master}
     \end{subfigure}
        \caption{Replica 159 Rq and Wq}
        \label{fig:159_rq_wq_master}
\end{figure}

\begin{figure}[!ht]
     \centering
     \begin{subfigure}[b]{0.45\textwidth}
         \centering
         \includegraphics[width=\textwidth]{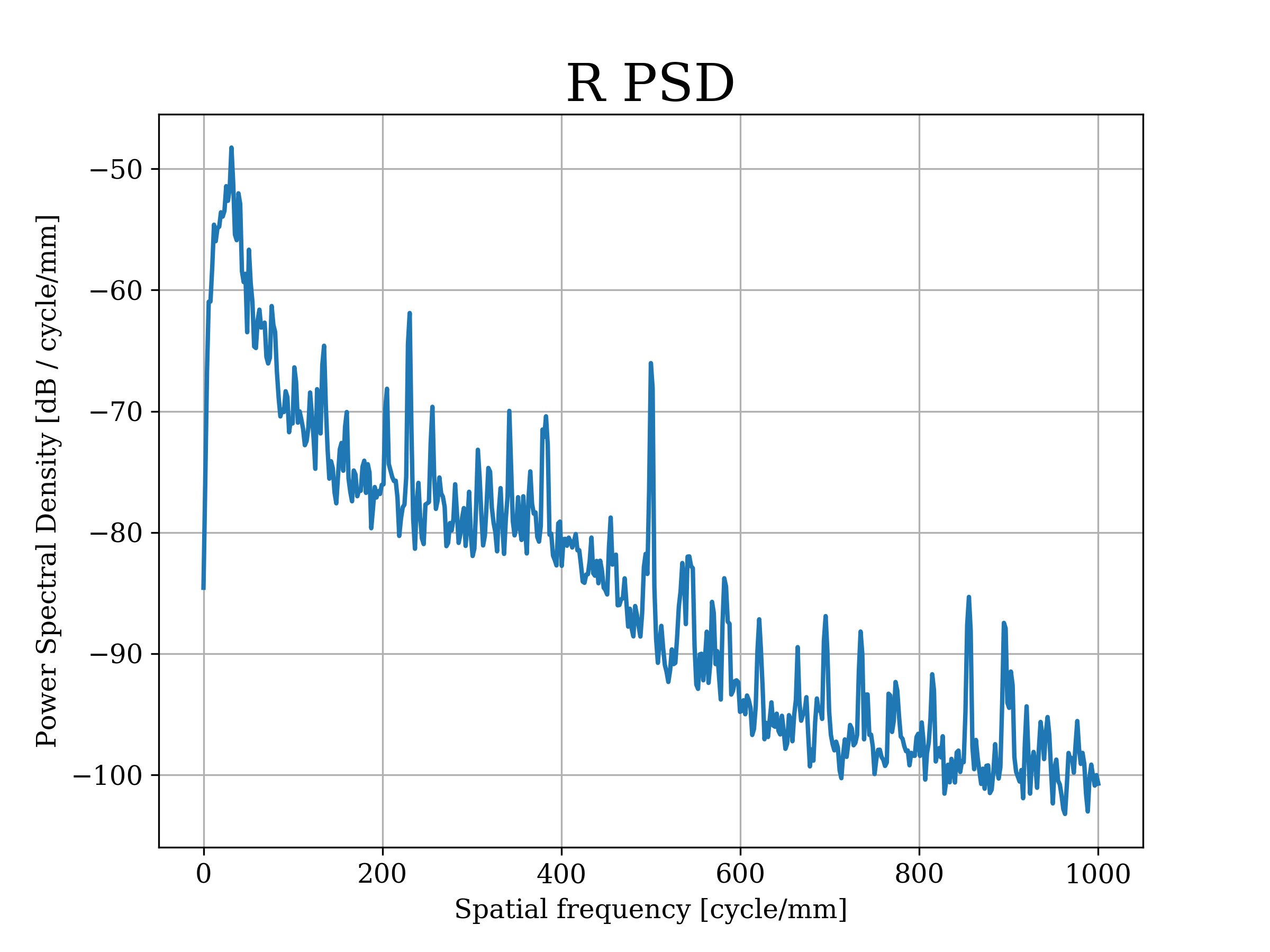}
         \caption{Rq Replica 172}
         \label{fig:172_r_master}
     \end{subfigure}
     \hfill
     \begin{subfigure}[b]{0.45\textwidth}
         \centering
         \includegraphics[width=\textwidth]{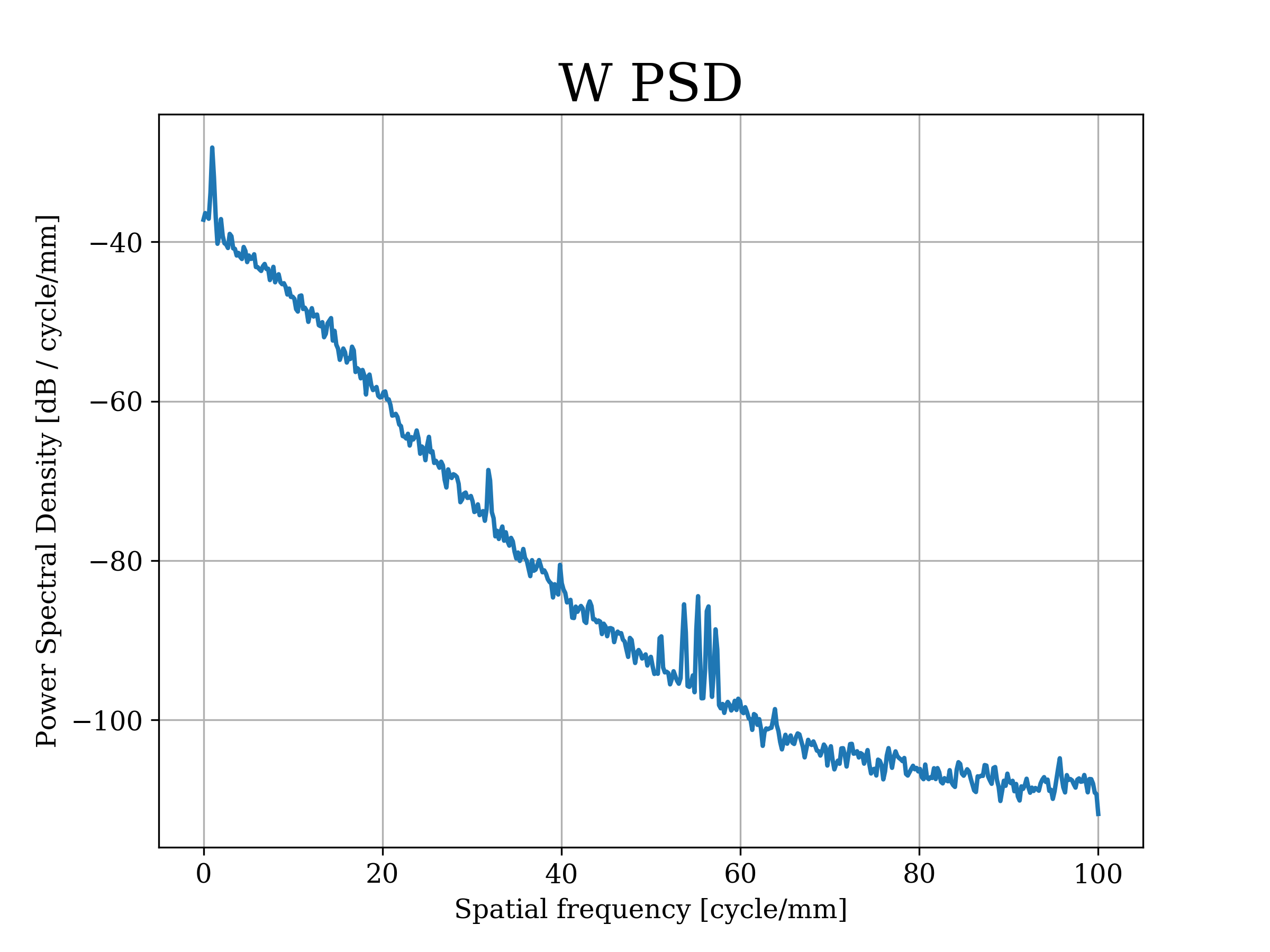}
         \caption{Wq Replica 172}
         \label{fig:172_w_master}
     \end{subfigure}
        \caption{Replica 172 Rq and Wq}
        \label{fig:172_rq_wq_master}
\end{figure}

\begin{figure}[!ht]
     \centering
     \begin{subfigure}[b]{0.45\textwidth}
         \centering
         \includegraphics[width=\textwidth]{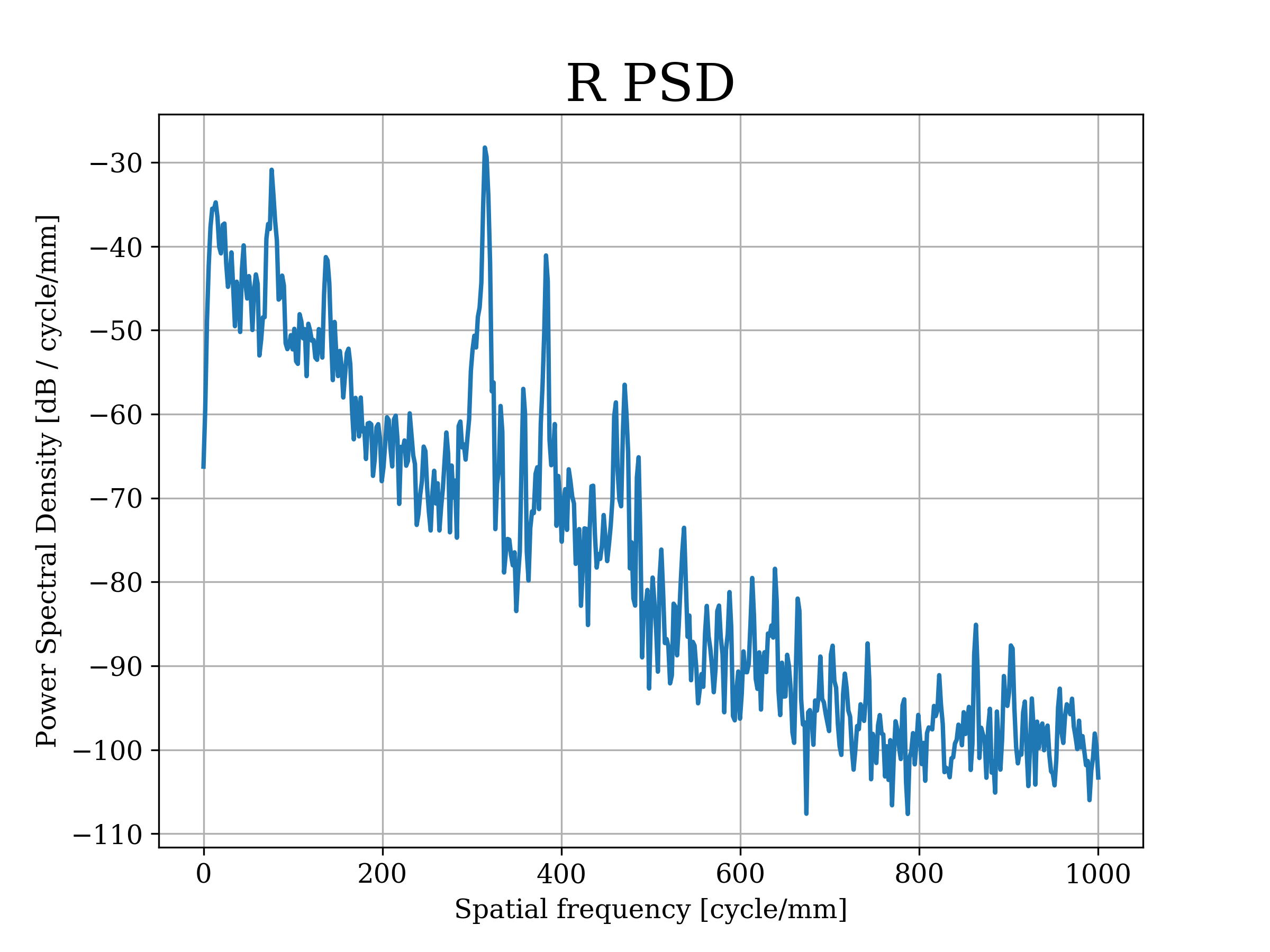}
         \caption{Rq Replica 173}
         \label{fig:173_r_master}
     \end{subfigure}
     \hfill
     \begin{subfigure}[b]{0.45\textwidth}
         \centering
         \includegraphics[width=\textwidth]{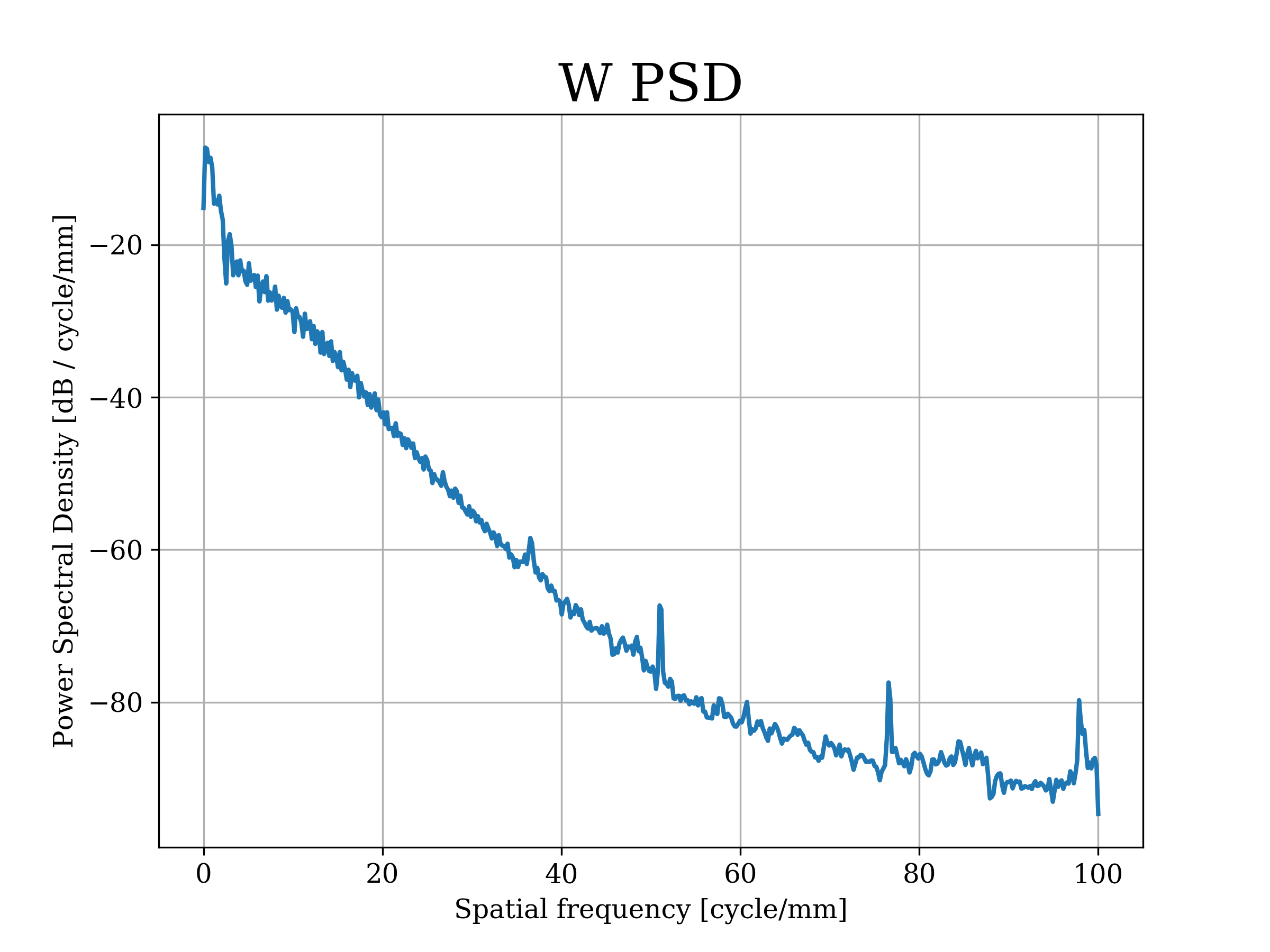}
         \caption{Wq Replica 173}
         \label{fig:173_w_master}
     \end{subfigure}
        \caption{Replica 173 Rq and Wq}
        \label{fig:173_rq_wq_master}
\end{figure}

\clearpage
\acknowledgments     

All the authors acknowledge financial support from Iniciativa Cient\'ifica Milenio v\'ia N\'ucleo Milenio de Formaci\'on Planetaria. A.B acknowledges support from FONDECYT grant 1190748, A.B. an N.S. acknowledges support from ESO Comit\'e-Mixto and A.B. from QUIMAL funding agencies. M.S., S.C and C.L acknowledge support from the ALMA-CONICYT fund. G.H and N.S acknowledge support from the Programa de Incentivo a la Iniciación Científica (PIIC) from USM.

\bibliography{report} 

\begin{thebibliography}{}
\makeatletter
\relax
\def\mn@urlcharsother{\let\do\@makeother \do\$\do\&\do\#\do\^\do\_\do\%\do\~}
\def\mn@doi{\begingroup\mn@urlcharsother \@ifnextchar [ {\mn@doi@}
  {\mn@doi@[]}}
\def\mn@doi@[#1]#2{\def\@tempa{#1}\ifx\@tempa\@empty \href
  {http://dx.doi.org/#2} {doi:#2}\else \href {http://dx.doi.org/#2} {#1}\fi
  \endgroup}
\def\mn@eprint#1#2{\mn@eprint@#1:#2::\@nil}
\def\mn@eprint@arXiv#1{\href {http://arxiv.org/abs/#1} {{\tt arXiv:#1}}}
\def\mn@eprint@dblp#1{\href {http://dblp.uni-trier.de/rec/bibtex/#1.xml}
  {dblp:#1}}
\def\mn@eprint@#1:#2:#3:#4\@nil{\def\@tempa {#1}\def\@tempb {#2}\def\@tempc
  {#3}\ifx \@tempc \@empty \let \@tempc \@tempb \let \@tempb \@tempa \fi \ifx
  \@tempb \@empty \def\@tempb {arXiv}\fi \@ifundefined
  {mn@eprint@\@tempb}{\@tempb:\@tempc}{\expandafter \expandafter \csname
  mn@eprint@\@tempb\endcsname \expandafter{\@tempc}}}

\bibitem[\protect\citeauthoryear{Ahmed, Tavakol, Das, Joven, Roozbehjavan  \&
  Minaie}{Ahmed et~al.}{2012}]{ahmed2012study}
Ahmed A.,  Tavakol B.,  Das R.,  Joven R.,  Roozbehjavan P.,   Minaie B.,
  2012, in SAMPE International Symposium Proceedings.

\bibitem[\protect\citeauthoryear{{Behrisch}}{{Behrisch}}{1981}]{Behrisch82}
{Behrisch} R.,  1981, {Sputtering by Particle Bombardment I}.
~"" Vol. 47, Springer-Verlag, \mn@doi{10.1007/3-540-10521-2}

\bibitem[\protect\citeauthoryear{{Hochhalter}, {Massarello}, {Maji}  \&
  {Fuierer}}{{Hochhalter} et~al.}{2006}]{hochhalter06}
{Hochhalter} J.~D.,  {Massarello} J.~J.,  {Maji} A.~K.,   {Fuierer} P.~A.,
  2006, in {Sasian} J.~M.,  {Turner} M.~G.,  eds,  Society of Photo-Optical
  Instrumentation Engineers (SPIE) Conference Series Vol. 6289, Society of
  Photo-Optical Instrumentation Engineers (SPIE) Conference Series. p. 628902,
  \mn@doi{10.1117/12.681042}

\bibitem[\protect\citeauthoryear{Hongkarnjanakul, Bouvet  \&
  Rivallant}{Hongkarnjanakul et~al.}{2013}]{HONGKARNJANAKUL2013549}
Hongkarnjanakul N.,  Bouvet C.,   Rivallant S.,  2013, \mn@doi [Composite
  Structures] {https://doi.org/10.1016/j.compstruct.2013.07.008}, 106, 549

\bibitem[\protect\citeauthoryear{Joyce}{Joyce}{2003}]{joyce2003common}
Joyce P.,  2003, United States Naval Academy, 1

\bibitem[\protect\citeauthoryear{{Monnier} et~al.,}{{Monnier}
  et~al.}{2018}]{Monnier18}
{Monnier} J.~D.,  et~al., 2018, \mn@doi [Experimental Astronomy]
  {10.1007/s10686-018-9594-1}, \href
  {https://ui.adsabs.harvard.edu/abs/2018ExA....46..517M} {46, 517}

\bibitem[\protect\citeauthoryear{Pham \& Marks}{Pham \& Marks}{2005}]{pham05}
Pham H.~Q.,  Marks M.~J.,  2005, Epoxy Resins.
American Cancer Society, p.~228 (\mn@eprint {}
  {https://onlinelibrary.wiley.com/doi/pdf/10.1002/14356007.a09\_547.pub2}),
  \mn@doi{https://doi.org/10.1002/14356007.a09_547.pub2}, \url
  {https://onlinelibrary.wiley.com/doi/abs/10.1002/14356007.a09\_547.pub2}

\bibitem[\protect\citeauthoryear{Schmidt}{Schmidt}{2008}]{schmidt08}
Schmidt J.,  2008, Papers from" UHON395 Carbonomics: Can Bioenergy save the
  World?", 1, 4

\bibitem[\protect\citeauthoryear{{Steeves}, {Laslandes}, {Pellegrino},
  {Redding}, {Bradford}, {Wallace}  \& {Barbee}}{{Steeves}
  et~al.}{2014}]{Steeves14}
{Steeves} J.,  {Laslandes} M.,  {Pellegrino} S.,  {Redding} D.,  {Bradford}
  S.~C.,  {Wallace} J.~K.,   {Barbee} T.,  2014, in {Navarro} R.,  {Cunningham}
  C.~R.,   {Barto} A.~A.,  eds,  Society of Photo-Optical Instrumentation
  Engineers (SPIE) Conference Series Vol. 9151, Advances in Optical and
  Mechanical Technologies for Telescopes and Instrumentation. p. 915105,
  \mn@doi{10.1117/12.2056560}

\bibitem[\protect\citeauthoryear{Wei, Zhang  \& Gong}{Wei et~al.}{2017}]{wei17}
Wei L.,  Zhang L.,   Gong X.,  2017, \mn@doi [Photonic Sensors]
  {10.1007/s13320-017-0388-2}, 7

\makeatother
\end{thebibliography}
\bibliographystyle{spiebib}

\end{document}